\documentclass[twocolumn,groupedaddress,amsmath,aps,longbibliography,nofootinbib]{revtex4-1}
\usepackage{hyperref}
\usepackage{graphicx}
\usepackage{amsmath,amsfonts}
\usepackage{dcolumn}
\usepackage{bm}
\usepackage{color}
\usepackage{multirow}
\usepackage{float}
\usepackage{url}
\usepackage{physics}
\usepackage{subfigure}
\usepackage[utf8]{inputenc}

\begin{document}

\title{Strong-coupling superconductivity of the Heusler-type compound ScAu$_2$Al: {\it Ab-initio} studies}
\author{Gabriel Kuderowicz}
\author{Bartlomiej Wiendlocha}
\email{wiendlocha@fis.agh.edu.pl}
\affiliation{AGH University of Krakow, Faculty of Physics and Applied Computer Science, Aleja Mickiewicza 30, 30-059 Krakow, Poland}

\date{\today}

\begin{abstract}
The ScAu$_2$Al superconducting Heusler-type compound was recently characterized
to have the highest critical temperature of $T_c = 5.12$~K and the strongest electron-phonon coupling among the Heusler family. 
In this work, the electronic structure, phonons, electron-phonon coupling, and superconductivity of ScAu$_2$Al are studied using \textit{ab initio} calculations. The spin-orbit coupling significantly changes the electronic structure removing the van Hove singularity from the vicinity of the Fermi level.
In the phonon spectrum, low frequency acoustic modes, additionally softened by the spin-orbit interaction, strongly couple with electrons, leading to the electron-phonon coupling constant $\lambda=1.25$, a record high among Heuslers.
The density functional theory for superconductors is then used to analyze superconducting state in this two-band superconductor. 
The effect of spin fluctuations (SF) on superconductivity is also analyzed.
The calculated critical temperatures of $T_c = 5.16$ K (4.79 K with SF) agree very well with the experiment, confirming the electron-phonon mechanism of superconductivity and showing a weak spin-fluctuations effect. 
The superconducting gaps formed on two Fermi surface sheets exhibit moderate anisotropy. Their magnitudes confirm the strong coupling regime, as the reduced average values are $2\Delta_{b_1}/k_BT_c \simeq 4.1$ and $2\Delta_{b_2}/k_BT_c \simeq 4.3$. Anisotropy of the gaps and large spread in their values significantly affect the calculated quasiparticle density of states.

\end{abstract}

\maketitle

\section{Introduction}
\begin{figure}[b]
	\centering
	\includegraphics[width=0.99\columnwidth]{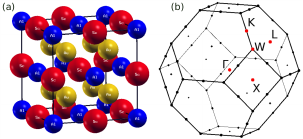}
	\caption{(a) Unit cell of the ScAu$_2$Al full Heusler compound with interpenetrating ${\it fcc}$ sublattices; (b) Brillouin zone with high-symmetry points.}\label{fig_cryst}
\end{figure}

Heusler family of compounds consists of more than a thousand systems with a wide variety of physical properties. To name a few, one can find half-metallic ferromagnetism \cite{Chen2006,Kourov2015,Mouatassime2021,Shan2009,Felser2015,Graf2011,Idrissi2020a,Idrissi2020b,Idrissi2020c,Idrissi2021}, shape memory effect \cite{Liu2003,Acet2012}, heavy fermion behavior \cite{Takayanagi1988,Lahiouel1987,Nakamura1988,Kaczorowski2003,Gofryk2005}, charge density waves \cite{Gruner2017}, topologically non-trivial states \cite{AlSawai2010,Liu2016,Sun2018,Pavlosiuk2016,PengJie2017} and superconductivity \cite{Wernick1983,Klimczuk2012,Pavlosiuk2016b,Pavlosiuk2015,Winterlik2009,Winiarski2021}.
In many cases, the properties of Heusler compounds can be easily tuned by chemical substitution \cite{Shan2009,Wollmann2017,Alijani2011,Graf2011b} or pressure \cite{Fukatani2011,Wang2016}. 
These intermetallic materials have a general formula XY$_2$Z for the so-called full Heusler compounds and XYZ for their half-Heusler counterparts. The full Heusler compounds crystallize in a cubic structure with a space group $Fm$-3$m$, No. 225, shown in Fig.~\ref{fig_cryst}. The relative simplicity of the structure and the wealth of physical properties make this family a very active and compelling field of research.

To our knowledge, to date superconductivity has been reported in 34 full Heusler compounds \cite{Winiarski2021}. Generally, they follow Matthias' rule, which states that the maximum superconducting transition temperature $T_c$ is expected for 5 to 7 valence electrons per atom in the unit cell \cite{Matthias1953,Matthias1955}. 
The highest $T_c$ among Heusler compounds was recently reported to be 5.12 K in ScAu$_2$Al \cite{Bag2022} which has 7 electrons per atom (28 per formula unit).
However, a considerably lower $T_c = 4.4$~K was previously reported for that system \cite{Poole2000,Benndorf2015,Winiarski2021}. Moreover, on the basis of the McMillan formula~\cite{McMillan1968}, measured Debye temperature ($\theta_D = 180$~K) and using a Coulomb pseudopotential value of $\mu^* = 0.13$ Bag {\it et al.}~\cite{Bag2022} estimated the electron-phonon coupling constant to be $\lambda=0.77$, suggesting a moderate strength of the electron-phonon interaction, not necessarily expected for a case with the highest $T_c$ in this family of materials. 

This raises the question of whether strongly coupled superconductivity with $\lambda > 1.0$ actually exists among Heuslers.
In other intermetallic superconducting families, such as Laves phases or A-15 compounds, materials with the highest $T_c$s exhibit $\lambda > 1.0$, for example, SrIr$_2$ ($\lambda = 1.1$, $T_c$ = 6.1~K)~\cite{gutowska2021}, CaIr$_2$ ($\lambda = 1.05$, $T_c = 7$ K)~\cite{haldo-cair2,cair2-carh2-tutuncu} or Nb$_3$Ge ($\lambda \simeq 1.7$, $T_c = 21.8$ K)~\cite{a15}. Therefore, it is worth considering whether the Heusler family is in some way ''distinct'' and incapable of achieving higher values of $\lambda$ than the postulated $\sim 0.8$, or the estimation of $\lambda$ based on the McMillan formula provides a very inaccurate result. These facts served as motivation for the undertaking of theoretical studies for ScAu$_2$Al.

This work provides a theoretical analysis of superconductivity in ScAu$_2$Al to determine the electron-phonon coupling strength, critical temperature, and the structure and anisotropy of the superconducting gap.
Calculations will help verify the strong- or moderate-coupling regime of superconductivity and to determine which of the experimentally reported values of $T_c$ corresponds to the one for the ideal ScAu$_2$Al phase, as due to the presence of defects and disorder, frequently observed in Heuslers, the discrepancy in experimental $T_c$'s is not unusual. 
Using the {\it ab initio} methods, we calculate the electronic structure, phonons, and electron-phonon coupling functions using density functional theory (DFT). To investigate the anisotropy of the superconducting gap and obtain the critical temperature without using any external parameters, the density functional theory for superconductors (SCDFT) in the decoupling approximation \cite{Oliveira1988,Luders2005,Kawamura2017,Kawamura2020} is used. 
In these calculations, we avoid the need to assume the value of the effective Coulomb repulsion parameter $\mu^*$ present in McMillan~\cite{McMillan1968} or Allen-Dynes \cite{Allen1975} $T_c$ formulas, and the ferromagnetic spin-fluctuations effect, competing with superconductivity, can be analyzed.

{The main results of our calculations show that ScAu$_2$Al is a two-band strong-coupled electron-phonon superconductor with $\lambda \sim 1.3$, significantly increased  by the spin-orbit coupling, $T_c \simeq 5$~K and $2\overline\Delta$/k$_BT_c \simeq 4.1$ with a moderate anisotropy of the two superconducting gaps.}

\section{Computational details}
DFT calculations were performed using the {\sc quantum espresso} package \cite{Giannozzi2009,Giannozzi2017}.
Projected augmented wave (PAW) pseudopotentials from the PSlibrary \cite{DalCorso2014,psnames} were used, with the Perdew-Burke-Ernzerhof exchange correlation functional \cite{Perdew1996}.
The effects of spin-orbit coupling (SOC) were studied by performing the scalar-relativistic and full-relativistic calculations.
Self-consistent calculations were done on a 12x12x12 Monkhorst-Pack $\mathbf{k}$-point mesh. 
Electronic density of states (DOS) and eigenvalues for electron-phonon matrix elements were obtained on a 24x24x24 grid, whereas the Fermi surface (FS) was rendered from a 48x48x48 grid. Kinetic energy cut-offs in the plane wave expansion of wavefunctions and charge density were set to 100 Ry and 600 Ry, respectively. In phonon calculations, dynamical matrices were calculated for a 6x6x6 $\mathbf{q}$-point grid which corresponds to 16 inequivalent {\bf q} points, from which the interatomic force constants in real space were calculated using Fourier interpolation methods. In the Supplemental Material~\cite{suppl} we show that these {\bf k-} and {\bf q-} point samplings were sufficient to obtain convergent results and discuss convergence against smearing parameters of the cold-smearing technique, used during integrations. In addition, a comparison of the electronic structure obtained in the pseudopotential calculations and from the full-potential all-electron linearized augmented plane wave method, implemented in the {\sc wien2k} package~\cite{wien2k,blaha2020}, is shown~\cite{suppl}, which validates the choice of pseudopotentials.

Superconducting density functional theory (SCDFT) calculations were carried out using the Superconducting Toolkit (SCTK) \cite{Kawamura2020} which is interfaced with {\sc quantum espresso}. 
As this procedure requires a shifted $\mathbf{q}$-point mesh, the prerequisite electron-phonon calculations were repeated on a shifted 6x6x6 $\mathbf{q}$-point grid, containing 28 independent points, with the 12x12x12 and 24x24x24 {\bf k-}point mesh for self-consistent cycle and eigenvalues calculations. Here, all integrations are done using the tetrahedron method.

The volume of the unit cell was relaxed with the Broyden-Fletcher-Goldfarb-Shanno algorithm. The obtained lattice constants, 6.5894 $\mathrm{\AA}$ without SOC and 6.5775 $\mathrm{\AA}$ with SOC, are slightly larger than the experimental one, equal to 6.5305 $\mathrm{\AA}$ \cite{Bag2022}. Such an overestimation is typical for the GGA exchange correlation functional.

\section{Results and discussion}
\subsection{Electronic structure}

\begin{figure}[b]
	\centering
	\includegraphics[width=0.99\columnwidth]{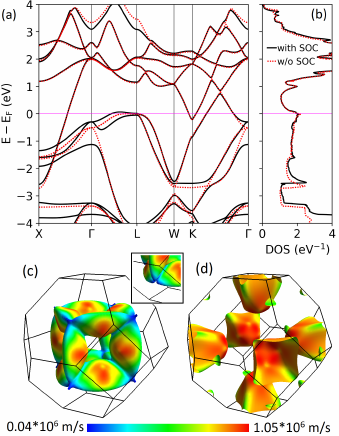}
	\caption{(a) Electronic bands and (b) total DOS of ScAu$_2$Al. (c), (d) Fermi surface with SOC and colored with the Fermi velocity. Inset shows a fragment near the L point calculated without SOC. FS was visualized in FermiSurfer \cite{Kawamura2019}.\label{fig_bandsfs}}	
\end{figure}

\begin{table}[b]
\caption{Calculated total and projected $N(E_F)$ of ScAu$_2$Al, and bandstructure values of the Sommerfeld coefficient $\gamma_{\rm band}$.\label{tab_eldos}}
\begin{center}
\begin{ruledtabular}
\begin{tabular}{cccccc}
         & \multicolumn{4}{c}{$N(E_F)$ (eV$^{-1}$)} & $\gamma_{\rm band}$\\
		 & total & Sc & Au$_2$ & Al & $\mathrm{\left(\frac{mJ}{mol\, K^2}\right)}$\\
		\hline
  		with SOC & 2.010 & 0.972 & 0.362 & 0.556 & 4.74\\
		w/o SOC & 2.043 & 0.990 & 0.359 & 0.565 & 4.82\\
	\end{tabular}
\end{ruledtabular}
\end{center}
\end{table}

\begin{figure*}[t]
	\centering
	\includegraphics[width=0.99\textwidth]{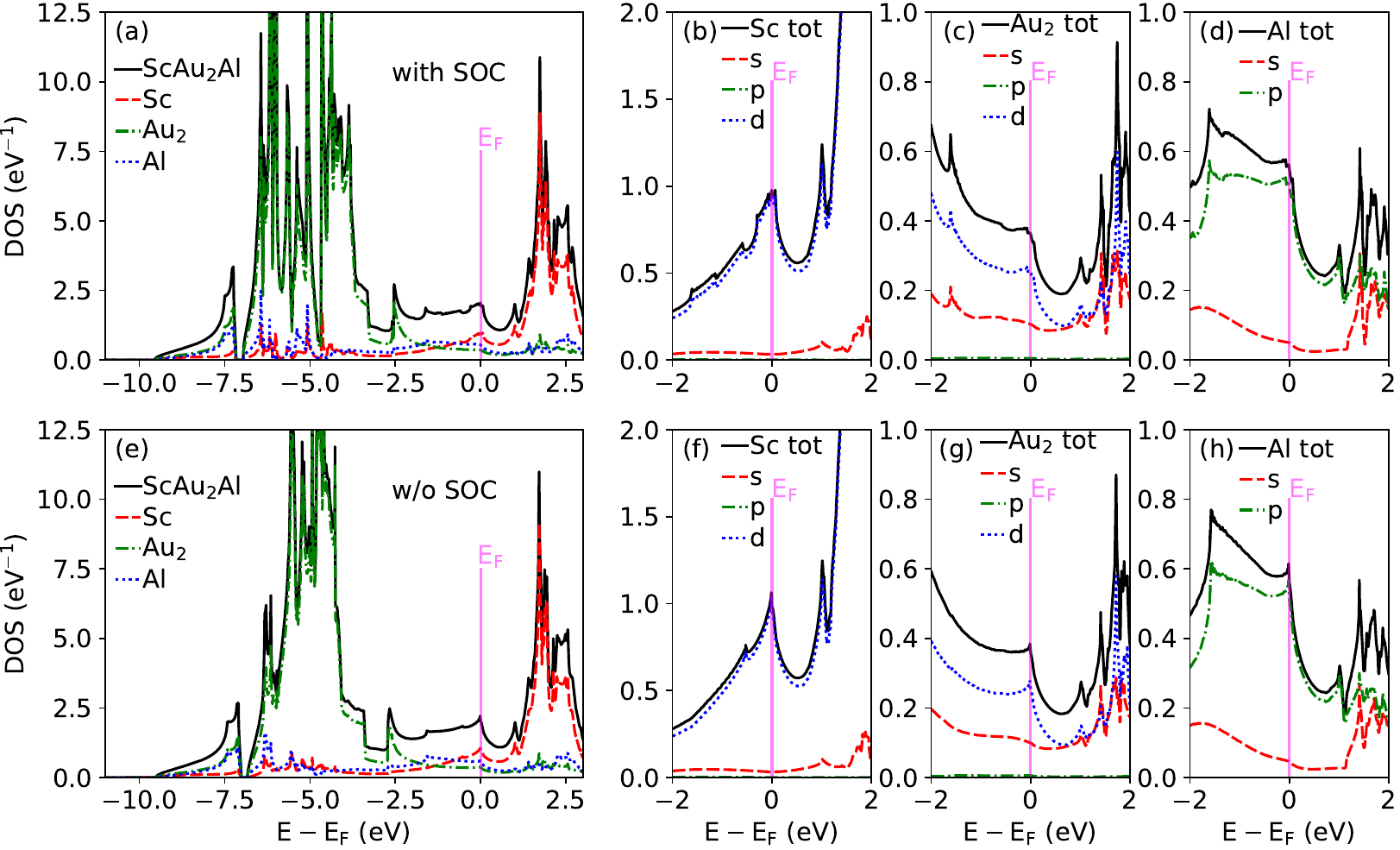}
	\caption{Total and projected density of states (DOS) of ScAu$_2$Al. Top panels show the full-relativistic results, bottom with SOC neglected. In panels (c) and (g) the densities of states are plotted for two Au atoms, the single-atomic orbital contribution is two times smaller.}\label{fig_eldos}
\end{figure*}

The calculated electronic dispersion relations and the Fermi surface of ScAu$_2$Al are shown in Fig.~\ref{fig_bandsfs}, and the densities of states (DOS) in Fig.~\ref{fig_eldos}. Brillouin zone with the high-symmetry points is shown in Fig~\ref{fig_cryst}(b).
In general, our results agree well with the earlier band-structure calculations \cite{Bag2022,SreenivasaReddy2015}. 

Two bands cross the Fermi level building two Fermi surface sheets [Fig.~\ref{fig_bandsfs}(c,d)]: one, hole-like, with a shape of a cube skeleton stretched between the L points, and the second, electron-like, with large pockets centered at X and small pockets around K. 
The first Fermi surface sheet, shown in Fig.~\ref{fig_bandsfs}(c), contributes approximately 75\% of the total density of states at the Fermi level, $N(E_F)$, as electrons have a lower Fermi velocity compared to the second sheet. 

The first notable feature in the electronic structure is a flat band near the L point, located at the Fermi energy $E_F$ in the scalar-relativistic case. Because the derivative $\nabla E(k)$ approaches zero, a van Hove singularity appears in the DOS as a sharp kink at $E_F$ in Fig.~\ref{fig_eldos}(e-h). 
The inset in Fig.~\ref{fig_bandsfs} shows a piece of the Fermi surface near the L point, calculated without SOC. As one can see, there are no states exactly at the L point and a closer analysis shows that, in fact, the electronic band is located 10 meV below $E_F$. 

Spin-orbit coupling visibly affects bands near the Fermi level in the $\Gamma$ -L direction, removing the band degeneracy. 
Two flat bands are moved away from the Fermi energy, removing the van Hove singularity from the vicinity of $E_F$. This results in the formation of two DOS peaks that merge into a broader maximum visible in Fig.~\ref{fig_eldos}(a-d), with slightly decreased $N(E_F)$, 
whereas in the Fermi surface small tubes around L are built, connecting the FS fragments [Fig.~\ref{fig_bandsfs}(c)].

\begin{figure}[t]
	\centering
	\includegraphics[width=0.99\columnwidth]{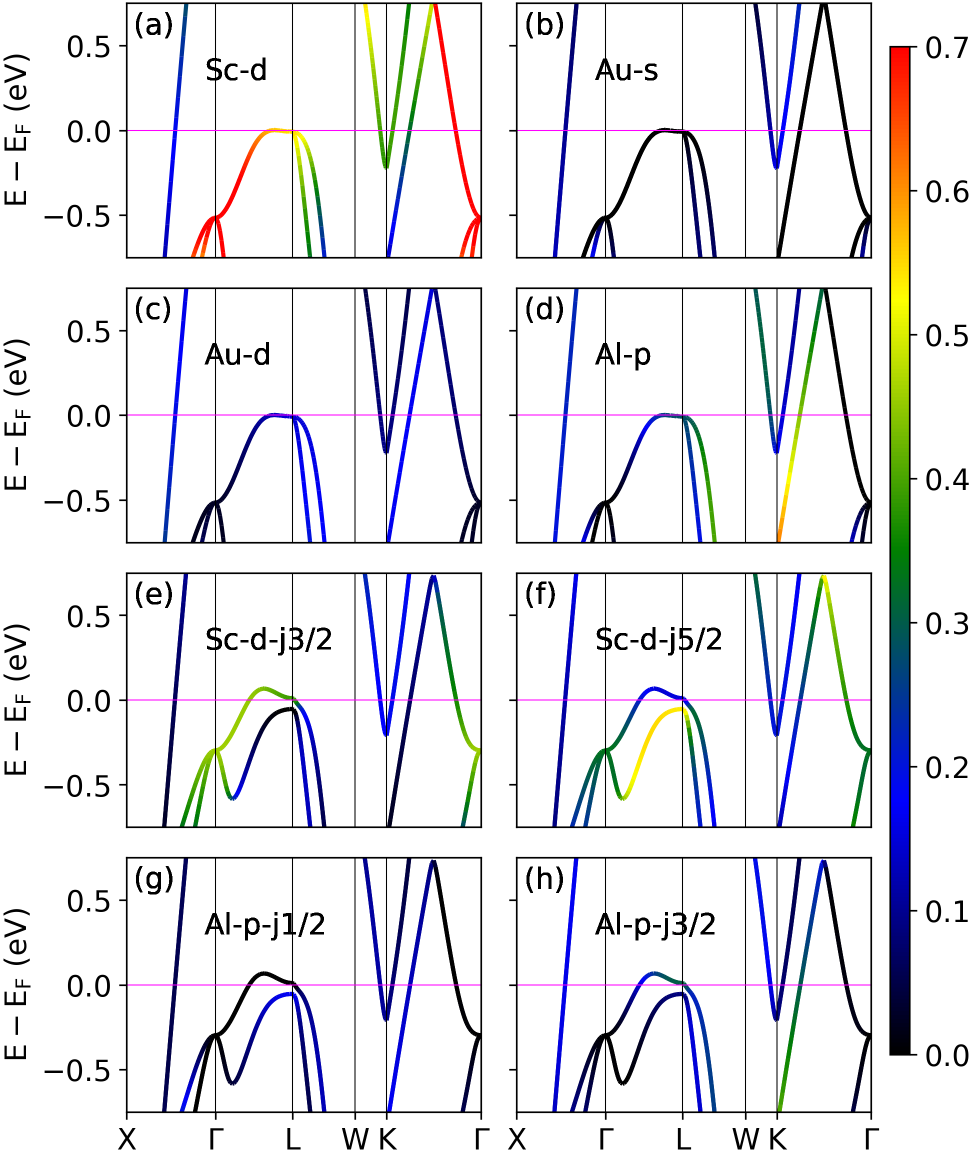}
	\caption{Electronic bands colored according to their orbital character. Panels (a-d) show the scalar-relativistic results, whereas in panels (e-h) spin-orbit coupling is included.}\label{fig_bandschar}	
\end{figure}

A more detailed view of the band structure is shown in Fig.~\ref{fig_bandschar} where the bands are colored according to their orbital character.

A dominant contribution of Sc-d states is noticeable, especially in the flat-band region and for a linear band between K and $\Gamma$. In the $\Gamma$ -L direction, SOC pushes Sc-d-$j_{5/2}$ below $E_F$, leaving the dominant contribution to Sc-d-$j_{3/2}$. The contributions from the Au-d and Al-p states, compared to Sc, are more uniformly distributed between the bands and as a function of the distance from $E_F$.

Analysis of densities of states in Fig.~\ref{fig_eldos} confirms that the states at $E_F$ are mainly built from Sc-3d, and the next contribution comes from Al-3p orbitals. 
Although the two Au atoms contribute 22 of 28 valence electrons in the ScAu$_2$Al unit cell, their almost filled 5d shells form bands deep below $E_F$. The total and projected values of $N(E_F$) are collected in Table~\ref{tab_eldos} and the largest contribution to DOS at the Fermi level comes from scandium.

Although spin-orbit coupling visibly affects the band structure near $E_F$, it has a small effect on the magnitude of $N(E_F)$. The relativistic value is only about 2\% smaller than the scalar-relativistic one.
The calculated density of states (with SOC) gives the electronic specific heat coefficient $\gamma_{\rm band} = 4.74$ mJ/(mol K$^2$). 
The experimental Sommerfeld coefficient, reported in~\cite{Bag2022} is $\gamma_{\rm expt} = 6.75$ mJ/(mol K$^2$).
That gives the electron phonon renormalization factor $\lambda_{\gamma}=\gamma_{\rm expt}/\gamma_{\rm band}-1$ = 0.425, which is very small considering the high $T_c$ = 5.12 K.
However, we consider the reported value of $\gamma_{\rm expt} = 6.75$ mJ/(mol K$^2$) as unlikely.
The value of $\gamma_{\rm expt}$ was obtained in Ref. \cite{Bag2022} by fitting the data of $C_p(T)$ measured in a magnetic field of 10 kOe  to the standard equation $C_p/T = \gamma + \beta T^2 + \delta T^4$. Measurements were reported for the temperature range 2~K - 5.5~K; the fit range was not reported.
In the Supplemental Material~\cite{suppl} we reanalyze the specific heat data from~\cite{Bag2022}, performing a set of fits in the temperature range 2.0 K - $T_{\rm max}$, with 4.5 K $<T_{\rm max}< 5.5$ K ($T_{\rm max}^2$ from 20 K$^2$ to 30 K$^2$). 
It appeared that the obtained $\gamma$ is very sensitive to the fit range; it varies from 12 mJ/(mol K$^2$) for the narrower fit range to 2.8 mJ/(mol K$^2$) for the broader, while it should be constant within some error bar. Similar situation holds for the Debye temperature, obtained from this fit.
The problem in stabilization of the temperature, mentioned in Ref.~\cite{Bag2022}, could have influenced the measurements; however,
as we discuss in the Supplemental Material~\cite{suppl} in temperatures above 2~K the lattice specific heat in ScAu$_2$Al cannot be analyzed using the Debye $C = \beta T^3 + \delta T^5$ formula, due to the presence of low-frequency phonon mode, discussed in the next paragraph. To precisely determine the Sommerfeld parameter and limit the influence of the lattice contribution, specific heat should be measured at lower temperatures, below 2 K.
Our theoretical value of the Sommerfeld parameter, renormalized by the calculated electron-phonon coupling constant (see the following sections), points to the value of $\gamma \sim 11$ mJ/(mol K$^2$).
Clearly, the specific heat in ScAu$_2$Al requires further experimental low-temperature studies to provide conclusive results and comparison.

\subsection{Phonons}

\begin{figure}[t]
	\centering
	\includegraphics[width=0.99\columnwidth]{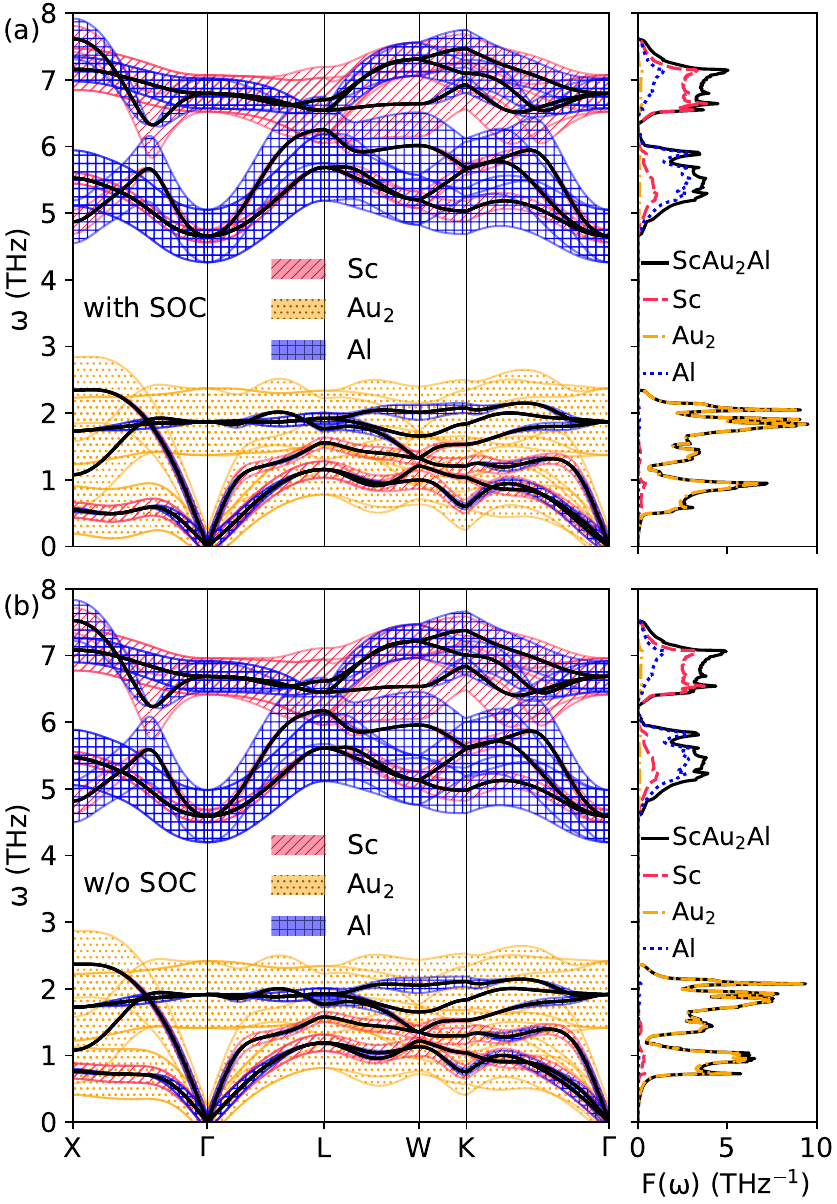}
	\caption{Phonon bands and partial phonon density of states of ScAu$_2$Al. Color shading with hatches represents atomic contributions to phonon branches.\label{fig_phbchar}}	
\end{figure}

\begin{figure}[t!]
	\centering
	\includegraphics[width=0.85\columnwidth]{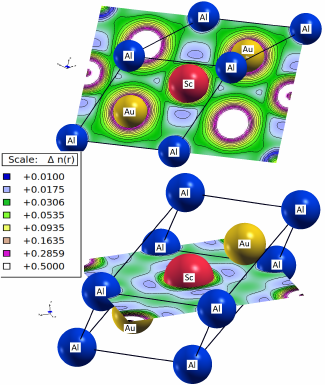}
	\caption{Electron density (number of electrons/$\rm a_B^3$, where $\rm a_B$ is the Bohr radius) in the primitive cell of ScAu$_2$Al. The electron density was visualized using XCrysDen \cite{Kokalj1999}.\label{fig_rho}}	
\end{figure}

Figure~\ref{fig_phbchar} presents the phonon dispersion relations $\omega_{\mathbf{q},\nu}$ and the phonon density of states F($\omega$). 
Gold, as expected for the heaviest atom, constitutes mostly the first six phonon modes with the lowest frequency, up to about 2.5 THz. These phonon branches have minor contributions from scandium and aluminum. 
A large gap in the phonon spectrum is formed, as the next phonons appear around 5 THz.
Such a separation of phonon modes is expected because Au atoms are approximately 8 and 4 times heavier than Al and Sc atoms, respectively.
The higher-frequency part of the phonon spectrum, involving Sc and Al vibrations, span the range from 5 THz to about 7.5 THz and is grouped into two parts. 
Three phonon modes, between roughly 5 THz and 6 THz, have a dominant Al character, whereas the highest three modes describe mostly the vibrations of Sc. 
Thus, inversion of the phonon modes is noticed, because despite the fact that Sc is 67\% heavier than Al, its phonon modes have much higher frequencies.
This inversion is caused by the stronger bonding of scandium atoms inside the structure, which leads to higher values of the force constants matrix elements. For example, the restoring force acting on Sc when it is displaced in the $x$ direction is about two times greater than for Al and Au, and the force constant matrix elements for the Sc-Au nearest-neighbors pair are approximately three times greater than those for the Al-Au pair.
This is supported by the charge density distribution, shown in 
Fig.~\ref{fig_rho} in the Al-Sc-Au and Sc-Al planes.  
The electron density between Sc and Au is generally larger than that between the Al and Au or Al and Sc pairs. This gives a larger metallic contribution to the mixed metallic-ionic bonding in this structure, 
strengthening the bonding between Sc and Au, compared to Al-Au or Al-Sc.  
In turn, higher values of the force constants result in higher frequencies of Sc vibrations, above Al.

\begin{figure}[b!]
	\centering
	\includegraphics[width=0.95\columnwidth]{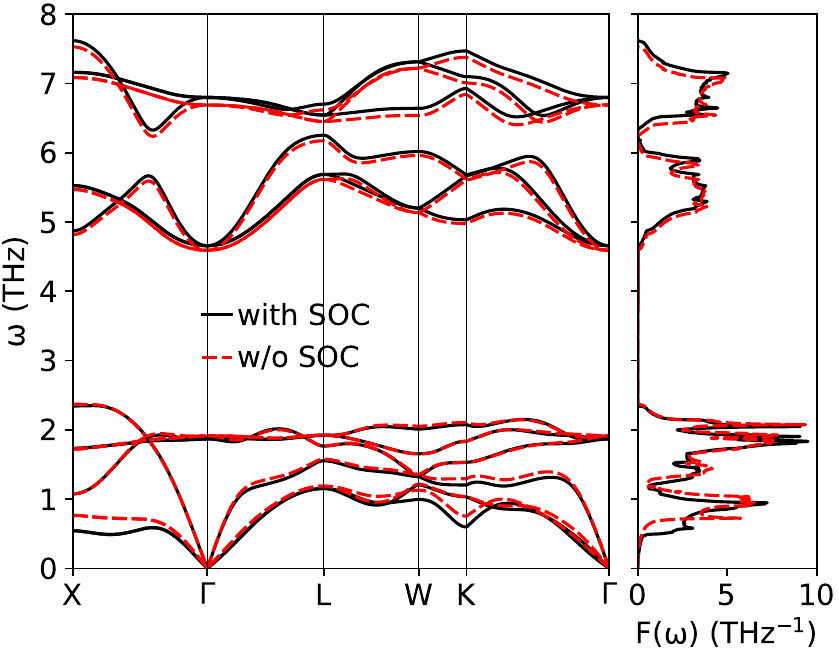}
	\caption{Effect of spin-orbit coupling on the phonon dispersion relations and phonon density of states of ScAu$_2$Al.\label{fig_phbands}}	
\end{figure}

Taking a closer look at the low frequency range, we notice a strong softening of the first two acoustic modes, especially in the $\Gamma$-X direction and at the K point. All three atom species make a visible contribution to these modes near the X point. On the other hand, the third acoustic mode at the X point has almost zero contribution from Sc and Al vibrations. SOC further enhances this softening as highlighted in Fig.~\ref{fig_phbands}, therefore, it must be included to accurately analyze superconductivity in ScAu$_2$Al, since the low-frequency phonon modes give the largest contribution to the electron-phonon coupling parameter $\lambda$. 
As far as the optical modes are concerned, the SOC slightly increases the phonon frequencies of the upper optical branches, which can be attributed to a slightly stiffened lattice because the lattice parameter computed with SOC is smaller than the scalar-relativistic one.
The average phonon frequencies are 3.82~THz (with SOC) and 3.80~THz (without SOC).

\begin{figure*}[t!]
	\centering
	\includegraphics[width=0.95\textwidth]{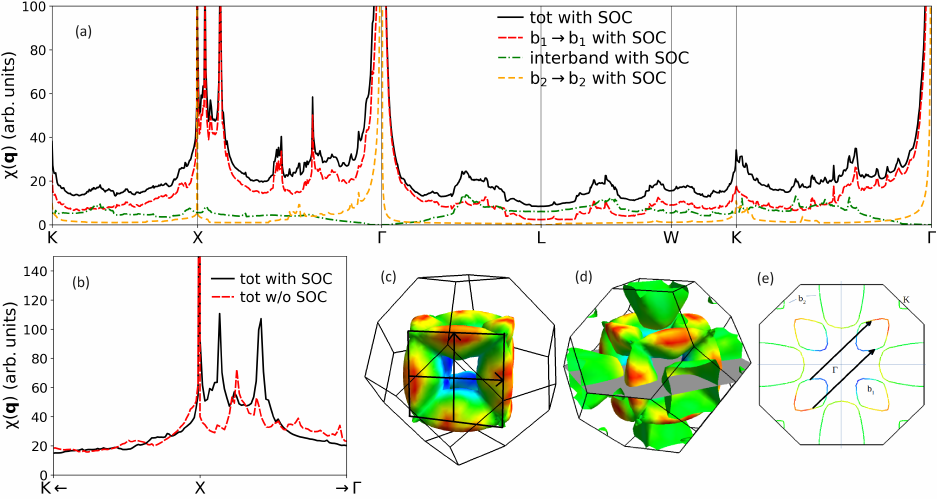}
	\caption{(a) Nesting function of ScAu$_2$Al with separated contributions from bands (see text), labels ''b$_1$'' and ''b$_2$'' refer to bands 1 and 2; (b) zoom near the X point; (c) nesting vectors parallel to $\Gamma$-X; (d) FS cut on which nesting vectors, parallel to $\Gamma$-K, are plotted in panel (e). Fermi surface is colored with electron-phonon coupling constant calculated using SCDFT (see Sect. \ref{sec:sctk}).} \label{fig_susc}	
\end{figure*}

One of the possible reasons for the softening of phonons in metals is the Kohn anomaly, which occurs if large fragments of the Fermi surface can be connected by a single vector $\mathbf{q}$ (FS exhibits strong nesting). In the case of ScAu$_2$Al the phonon branches in $\Gamma$-X gradually flatten, so there are no sharp dips in the dispersion relation, indicating a characteristic discontinuity of a derivative of $\omega(\mathbf{q})$. 
To quantitatively investigate the nesting of the Fermi surface, we calculated the values of the nesting function
\begin{align}
    \chi(\mathbf{q}) &= \sum_{n,n'} \int_{BZ} \frac{d^3\mathbf{k}}{V_{BZ}} \delta(E_F-E_{n,\mathbf{k}})\delta(E_F-E_{n',\mathbf{k}+\mathbf{q}}).
\end{align}
Figure~\ref{fig_susc}(a) shows $\chi(\mathbf{q})$ calculated in selected directions in the Brillouin zone. 
The largest number of states are connected by vectors starting from the first band, which builds the FS sheet shown in Fig.\ref{fig_bandsfs}(c) and has large contributions from Sc-d-$j_{3/2}$. 
Apart from the irrelevant peak in $\Gamma$, the nesting function is strongly enhanced in the $\Gamma$-X direction and near the X point, which is correlated with the observed softening of the first acoustic phonon mode. Furthermore, $\chi(\mathbf{q})$ is enhanced at the K point, where we also noticed softened phonons.
Nesting vectors $\Gamma$-X and $\Gamma$-K are visualized on the Fermi surface in \ref{fig_susc}(c) and \ref{fig_susc}(e), respectively.
Furthermore, spin-orbit coupling significantly increases $\chi(\mathbf{q})$ in the $\Gamma$-X direction, as shown in Fig.~\ref{fig_susc}(a,b), which partly answers the question why the $\Gamma$-X acoustic mode is additionally softened in the relativistic case. 
Analysis of $\chi(\mathbf{q})$ shows that the specific geometry of the Fermi surface is beneficial for strong electron-phonon coupling and superconductivity in ScAu$_2$Al.

\begin{figure}[t]
	\centering
	\includegraphics[width=0.99\columnwidth]{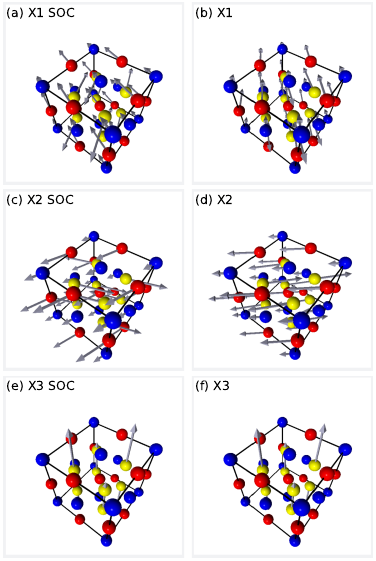}
	\caption{Acoustic phonon modes at X=(0,1,0) in the unit cell of ScAu$_2$Al. Sc, Au and Al atoms are colored red, yellow and blue, respectively.\label{fig_vibplotX}}	
\end{figure}

To further investigate the character of the softened phonon mode in $\Gamma$ -X, we visualized the movement of the atoms in the real space for the phonons at the X point in Fig.~\ref{fig_vibplotX}. 
This also allows one to see the additional difference imposed by spin-orbit coupling, as SOC increases the $\chi(\mathbf{q})$ values in  the $\Gamma$-X direction, but also modifies the atomic displacement, leading to a much more softened phonon mode in the fully-relativistic case, compared to the scalar-relativistic one.
What we can observe is that without SOC in the first two degenerate modes, every atom moves along the diagonal of the base of the unit cell, in the direction where all atoms are aligned.
Spin-orbit coupling by influencing the electronic structure, charge densities, and force constants,
qualitatively changes this movement pattern because then both Au atoms vibrate in a direction at an angle approximately 130$^\circ$ to the one of Sc and Al, in a less dense-packed direction. This results in a lowering of the vibration frequency.
The last acoustic mode at the X points consists only of Au vibrations, and neither its frequency nor the normal mode eigenvectors are visibly affected by the SOC.
In the Supplemental Material~\cite{suppl} animations of the lowest-frequency phonon modes at X, W, K, L and the middle of X-$\rm \Gamma$ are additionally provided. It is interesting to note that at the W point Au atoms move in circles, so the investigation of the chirality of phonons~\cite{chiral-prl,chiral-abi,chiral-abi2} in ScAu$_2$Al may be potentially interesting. 
Furthermore, SOC also causes a circular trajectory of Au movement in the point in the middle of X-$\rm \Gamma$, further highlighting that SOC substantially changes the vibrational properties of ScAu$_2$Al.
Chirlal phonon effects would be especially interesting if it were possible to break the time-reversal and inversion symmetries of the structure by synthesizing a noncentrosymmetric half-Heusler analog with magnetic dopants. In such a case an overall phonon angular momentum, carried by individual chiral modes, does not vanish, and this could be a possible interesting point for further studies on related Heusler structures. 

\subsection{Electron-phonon coupling}

Electron-phonon interaction is described by the hamiltonian \cite{Grimvall1981,Giustino2017,Wierzbowska2005}
\begin{equation}
    \hat{H} = \sum_{\mathbf{k},\mathbf{q},\nu,i,j} g_{\mathbf{q}}^{\nu}(\mathbf{k},i,j) c_{\mathbf{k}+\mathbf{q}}^{\dagger i} c_{\mathbf{k}}^{j} (b_{-\mathbf{q}}^{\dagger\nu}+b_{\mathbf{q}}^{\nu}).
\end{equation}
Operators $c$ refer to electrons and $b$ to phonons, with lower indices describing wave vectors and upper band numbers ($\mathbf{q}$ and $\nu$ for phonons and $\mathbf{k}, i, j$, for electrons). The matrix element $g_{\mathbf{q}}^{\nu}(\mathbf{k},i,j)$ has a form \cite{Wierzbowska2005}
\begin{equation}
 g_{\mathbf{q}}^\nu(\mathbf{k},i,j)=\sum_s \left ( \frac{\hbar}{2M_s\omega_{\mathbf{q}\nu}} \right ) ^{1/2} \langle \psi_{i,\mathbf{k}}| \frac{dV_{scf}}{d\hat{u}_{\nu,s}} \cdot \hat{\epsilon}_\nu | \psi _{j,\mathbf{k+q}} \rangle,
\end{equation}
where index $s$ labels atoms in the unit cell, $\psi_{i,\mathbf{k}}$ is an electronic wavefunction, $\frac{dV_{scf}}{d\hat{u}_{\nu,s}}$ is a self-consistently calculated change in the electronic potential caused by displacing atom in the direction $\hat{u}_{\nu,s}$ and $\hat{\epsilon}_\nu$ is a phonon polarization vector. Then one can introduce the phonon linewidth~\cite{allen-linewidths}
\begin{align}
	\gamma_{\mathbf{q}}^{\nu} = 2\pi \omega_{\mathbf{q}\nu} \sum_{ij} \int& \frac{d^3k}{\Omega_{BZ}}  |g_{\mathbf{q}}^{\nu}(\mathbf{k},i,j)|^2 \nonumber \\
\times& \delta(E_{\mathbf{q}}^{i} - E_F) \delta(E_{\mathbf{k}+\mathbf{q}}^{j} - E_F). \label{eq_gammaq}
\end{align}
It is worth noting that phonon linewidths do not directly depend on the phonon frequencies, thus they play a role of electronic contributions to the electron-phonon interaction.
Dirac deltas select only electrons from the Fermi surface to participate in the electron-phonon interaction, and the matrix element $g$ is the coupling strength. The phonon linewidth is inversely proportional to the phonon's lifetime. Now one can define the Eliashberg function
\begin{equation}
	\alpha^2F(\omega) = \frac{1}{2\pi N(E_F)} \sum_{\mathbf{q}\nu} \delta(\omega - \omega_{\mathbf{q}\nu}) \frac{\gamma_{\mathbf{q}\nu}}{\hbar \omega_{\mathbf{q}\nu}} \label{eq_a2F}
\end{equation}
and the electron-phonon coupling constant
\begin{equation}
	\lambda = 2 \int_0^{\omega_{max}} \frac{\alpha^2F(\omega)}{\omega} d\omega. \label{eq_lambda_a2F}
\end{equation}
A more detailed discussion can be found in \cite{Grimvall1981}.

\begin{figure}[t]
	\centering
	\includegraphics[width=0.99\columnwidth]{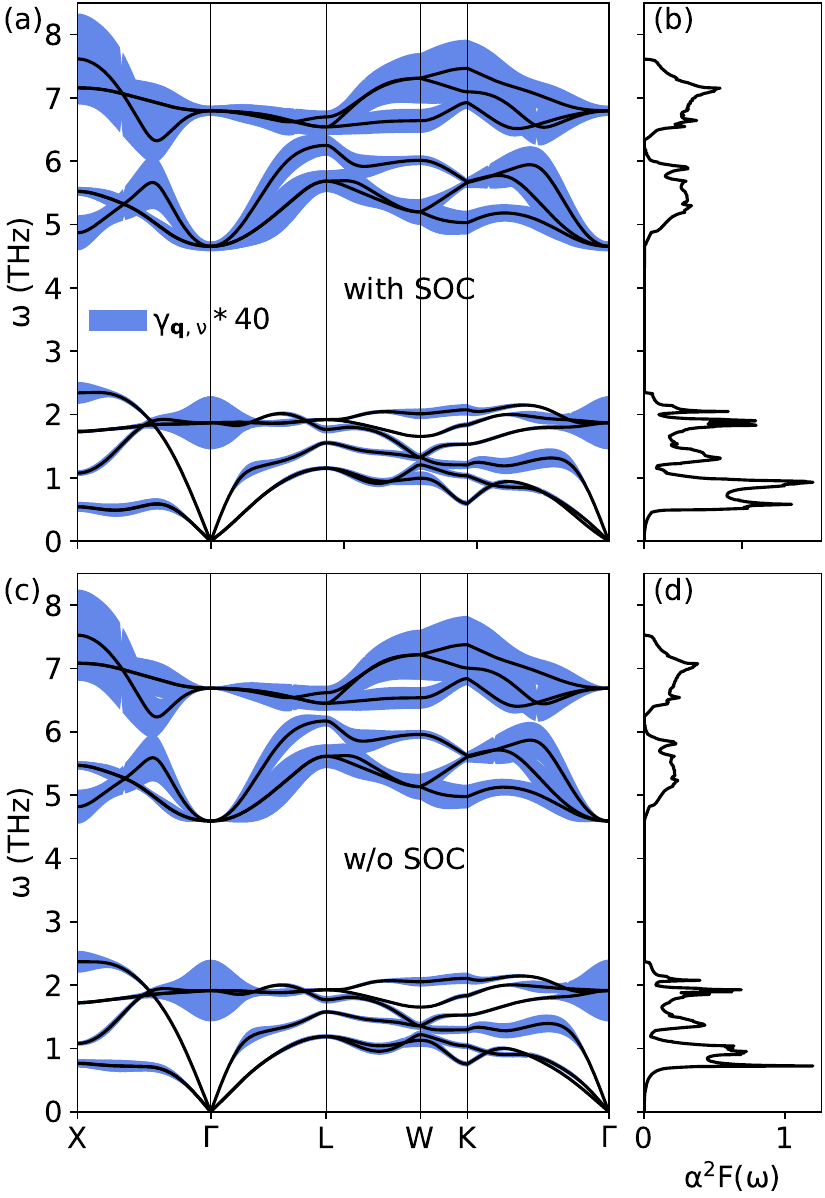}
	\caption{Phonon linewidths multiplied by 40 and Eliashberg functions of ScAu$_2$Al.\label{fig_gamlines}}	
\end{figure}

\begin{figure}[t]
	\centering
	\includegraphics[width=0.99\columnwidth]{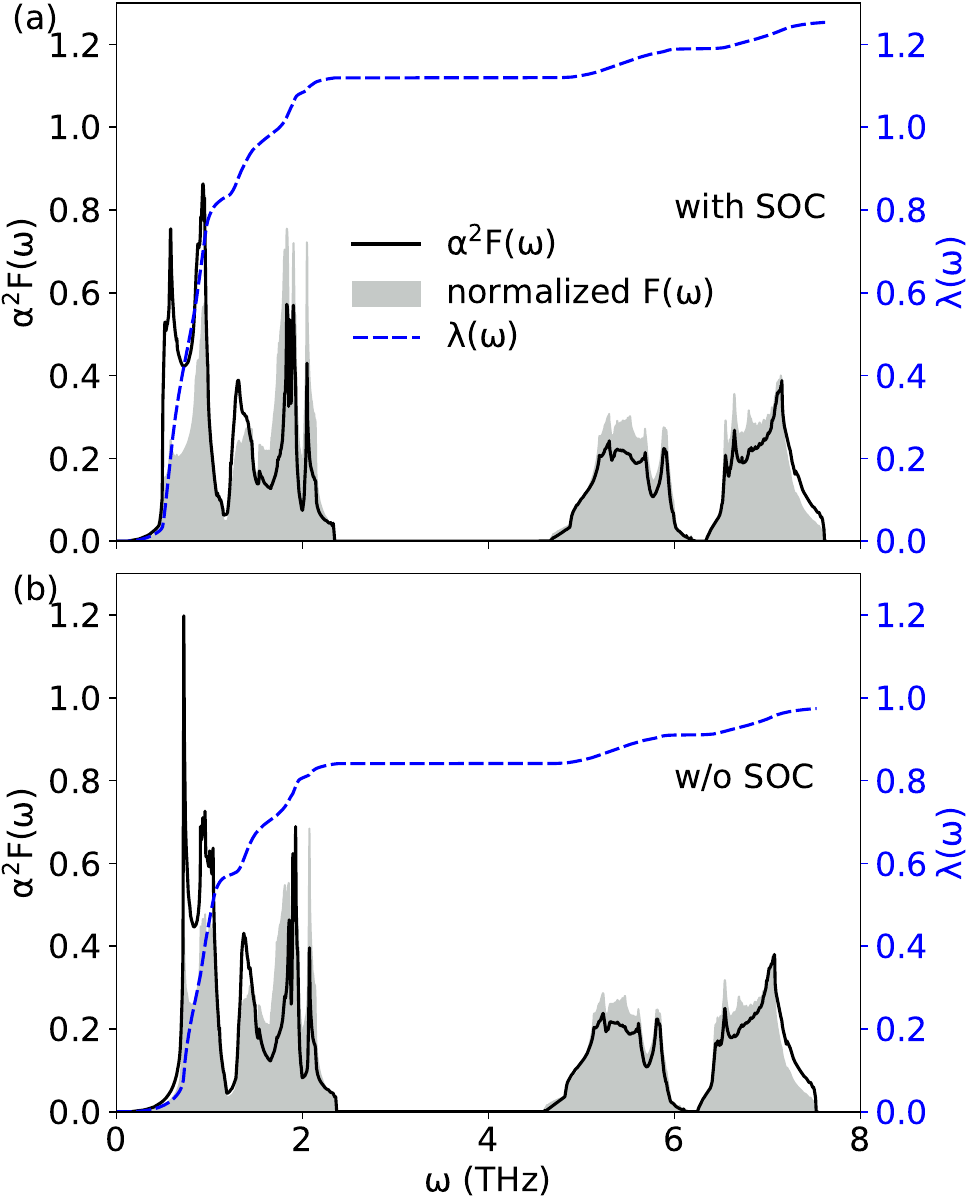}
	\caption{Eliashberg functions, normalized phonon density of states and cumulative electron-phonon coupling constant.\label{fig_a2F}}	
\end{figure}

Figure~\ref{fig_gamlines} shows the phonon linewidths plotted over the dispersion relations and the Eliashberg function. Typically, $\gamma_{\mathbf{q}}^{\nu}$ is small for acoustic modes, but here it is pronounced in the softened branch in the $\Gamma$-X direction. 
Because the Eliashberg function is inversely proportional to the phonon frequency, the contribution of the lowest mode enhances $\alpha^2F(\omega)$. It is well seen in Fig.~\ref{fig_a2F} where the Eliashberg function is plotted together with the phonon density of states $F(\omega)$. 
Furthermore, since the electron-phonon coupling constant $\lambda$ is proportional to $\alpha^2F(\omega)/{\omega}$, the contribution of the softened mode will be the largest. The cumulative frequency distribution of $\lambda(\omega)$, calculated by integrating equation (\ref{eq_lambda_a2F}) up to the frequency $\omega$, is shown in Fig.~\ref{fig_a2F}. 
The dominant role of the low-frequency Au-dominated modes (below 2.33~THz) in contributing to the electron-phonon coupling in ScAu$_2$Al is very well seen. They contribute approximately 86\% of $\lambda$ in the scalar-relativistic case, and the spin-orbit coupling increases this to 89\%. 

The scalar-relativistic electron-phonon coupling constant is $\lambda = 0.97$ and is remarkably enhanced in about 30\%, to $\lambda = 1.25$, when spin-orbit coupling is included. 
The contributions $\lambda_{\nu}$ from the each of the phonon modes ${\nu}$ are collected in Table~\ref{tab_lambx}. 
Analyzing these values, one can conclude that the relativistic increase of $\lambda$ is mostly contributed by the two lowest acoustic phonon modes, 
as $\lambda_1$ increases from 0.295 to 0.453 and $\lambda_2$ from 0.266 to 0.368.

To analyze whether this increase of $\lambda$ due to SOC is more related to the increase in the interaction strength (via the phonon linewidths) or due to decrease of phonon frequencies, we have calculated values of two integrals
\begin{equation}\label{eq:I}
    I = \int_0^{\omega_{\rm max}} \omega \cdot \alpha^2F(\omega) d\omega,
\end{equation}

\begin{equation}\label{eq:w2}
    \langle \omega^2 \rangle = \int_0^{\omega_{\rm max}} \omega \cdot \alpha^2F(\omega) d\omega
    \left/ \int_0^{\omega_{\rm max}} \alpha^2F(\omega) \frac{d\omega}{\omega}\right. .
\end{equation}

The integral $I$ is a frequency-independent measure of  electronic contribution to $\lambda$~\cite{gutowska2021,kuderowicz2022}, and $\langle \omega^2 \rangle$ is defined in such a way that $\lambda = \frac{2I}{\langle \omega^2 \rangle}$\footnote{In a monoatomic system the McMillan-Hopfield parameter~\cite{McMillan1968} is defined as $\eta = 2MI$ ($M$ is the atomic mass) and then the electron-phonon coupling constant is calculated using the well-known formula $\lambda = \frac{\eta}{M\langle \omega^2 \rangle}$}.
For the whole phonon spectrum, the values are $I=3.229$~THz$^2$, $\langle \omega^2 \rangle=5.153$~THz$^2$ (with SOC) and $I=3.122$~THz$^2$, $\langle \omega^2 \rangle=6.411$~THz$^2$ (no SOC), thus
the main reason for the increase in $\lambda$ due to SOC is the decrease of the phonon frequencies. 
Similar integrals computed for each phonon mode are collected in Table~\ref{tab_lambx}.

The calculated $\lambda$ places ScAu$_2$Al in the strong coupling regime and is the largest among the values reported in Heusler compounds (YPd$_2$Sn has $\lambda$ = 0.99 \cite{Tutuncu2014}).

\begin{table*}[t]
\caption{Parameters of the electron-phonon interaction in ScAu$_2$Al for the total phonon spectrum and individual phonon modes: the electron-phonon coupling parameter $\lambda$ [eq. (\ref{eq_lambda_a2F})], the frequency-independent electronic contribution $I$ [eq. (\ref{eq:I})] and ,,average square frequency'' $\langle\omega^2\rangle$ [eq. (\ref{eq:w2})].\label{tab_lambx}}
\begin{center}
\begin{ruledtabular}
\begin{tabular}{cccccccccccccc}
 & total  & \multicolumn{12}{c}{value for the phonon mode no.} \\
&   & 1  & 2  & 3  & 4  & 5  & 6  & 7  & 8  & 9  & 10  & 11  & 12 \\
 $\lambda$ \\ %& total  & 1  & 2  & 3  & 4  & 5  & 6  & 7  & 8  & 9  & 10  & 11  & 12 \\
with SOC & 1.253 & 0.453 & 0.368 & 0.143 & 0.059 & 0.049 & 0.047 & 0.026 & 0.023 & 0.022 & 0.018 & 0.021 & 0.025\\
w/o SOC & 0.974 & 0.295 & 0.266 & 0.128 & 0.056 & 0.048 & 0.047 & 0.025 & 0.023 & 0.021 & 0.017 & 0.021 & 0.025\\
%\hline
$I$ (THz$^2$)& \\
 with SOC & 3.229 & 0.118 & 0.111 & 0.130 & 0.087 & 0.085 & 0.102 & 0.346 & 0.333 & 0.362 & 0.396 & 0.510 & 0.649 \\
 w/o SOC & 3.122 & 0.103 & 0.106 & 0.124 & 0.084 & 0.084 & 0.102 & 0.336 & 0.326 & 0.345 & 0.376 & 0.498 & 0.638 \\
%\hline
$\langle\omega^2\rangle$ (THz$^2$)& \\
 with SOC & 5.153 & 0.520 & 0.605 & 1.815 & 2.949 & 3.466 & 4.287 & 27.081 & 29.294 & 33.060 & 44.624 & 48.701 & 51.781 \\
 w/o SOC & 6.411 & 0.701 & 0.794 & 1.939 & 2.989 & 3.506 & 4.349 & 26.431 & 28.592 & 32.365 & 43.397 & 47.482 & 50.522 \\
	\end{tabular}
\end{ruledtabular}
\end{center}
\end{table*}

\begin{table}[b]
	\caption{Logarithmic average $\omega_{\rm ln}$, electron-phonon coupling constant $\lambda$, superconducting transition 
temperature $T_c$ from Allen-Dynes formula (\ref{eq_Allen_Dynes}) with $\mu^*=0.1$, and Sommerfeld coefficient $\gamma$ renormalized with 
$\lambda$ from the electron-phonon 
calculations, $\gamma = \gamma_{band}(1+\lambda)$. Experimental value of $T_c = 5.12$~K. \label{tab_lambdaTc}}
\begin{center}
\begin{ruledtabular}
\begin{tabular}{ccccc}
		 & $\omega_{\rm ln}$ & $\lambda$ & $T_c$ & $\gamma$\\
		 & (THz) & & (K)  & $\mathrm{\left(\frac{mJ}{mol\, K^2}\right)}$\\ %$\mathrm{(mJ \ mol^{-1} \ K^{-2})}$\\
		\hline
  		with SOC & 1.077 & 1.253 & 5.43 & 10.68\\
		w/o SOC & 1.324 & 0.974 & 4.57 & 9.51\\
	\end{tabular}
\end{ruledtabular}
\end{center}
\end{table}

The superconducting transition temperature may now be calculated from the Allen-Dynes formula~\cite{Allen1975}:
\begin{equation}\label{eq_Allen_Dynes}
k_{B}T_{c}=\frac{f_1 f_2 \hbar\omega_{\rm ln}}{1.20}\,
\exp\left\{-\frac{1.04(1+\lambda)}{\lambda-\mu^{*}(1+0.62\lambda)}\right\}.
\end{equation}
As $\lambda$ is on the order of 1.0, the strong-coupling prefactors $f_1$ and $f_2$
must be used.
They are defined as~\cite{Allen1975}:
\begin{equation}
f_1 = [1+(\lambda/\Lambda_1)^{3/2}]^{1/3},
\end{equation}
\begin{equation}
f_2 = 1 + \frac{\left( \bar{\omega_2}/\omega_{\rm ln} -1 \right) \lambda^2}{\lambda^2 + \Lambda_2^2},\ \bar{\omega_2} = \sqrt{2\langle \omega^2 \rangle},
\end{equation}
where
\begin{equation}
\Lambda_1 = 2.46(1+3.8\mu^*),
\end{equation}
and
\begin{equation}
\Lambda_2 = 1.82(1+6.3\mu^*)\left( \bar{\omega_2}/\omega_{\rm ln} \right).
\end{equation}
The logarithmic average phonon frequency is in a form:
\begin{equation}
	\omega_{\rm ln} = \exp \left( \frac{2}{\lambda}\int_0^{\omega_{max}} \alpha^2F(\omega) \ln \omega \frac{d\omega}{\omega} %\middle/ 
%\int_0^{\omega_{max}}\alpha^2F(\omega) \frac{d\omega}{\omega}  
\right).
\label{eq_omloga2F}
\end{equation}

\begin{figure*}[t]
	\centering
	\includegraphics[width=0.99\textwidth]{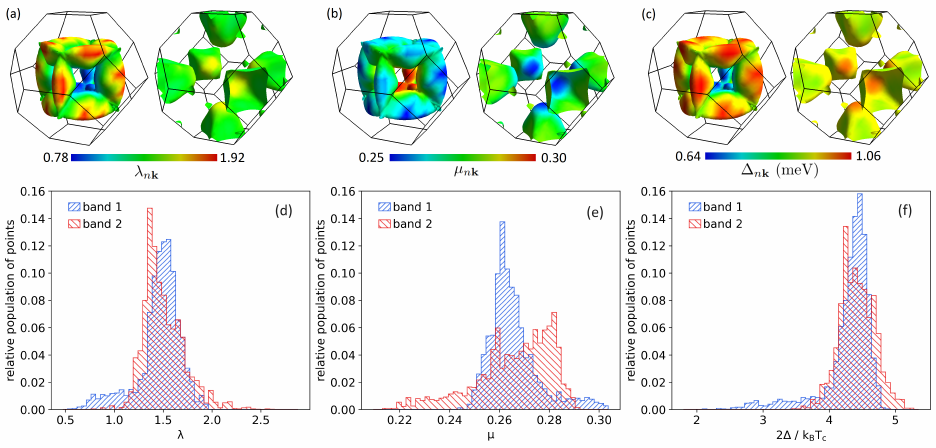}
	\caption{The $n\mathbf{k}$ dependent (a) electron-phonon coupling constant, (b) Coulomb potential and (c) superconducting gap function at T=0~K. Histograms in (d), (e), and (f) present the distribution of the quantities that were plotted on the Fermi surface. These results were computed with SOC and w/o SF.\label{fig_sctk}}	
\end{figure*}

The calculated $T_c$, $\lambda$ and $\omega_{\rm ln}$ are collected in Table~\ref{tab_lambdaTc}. Although the phonon modes reach approximately 7.5 THz, a small $\omega_{\rm ln} = 1.08$ THz (51.8 K, with SOC) is obtained. This is a consequence of
the enhancement of the Eliashberg function in the low-frequency range, especially in the fully relativistic case. The scalar-relativistic value of $\omega_{\rm ln} = 1.32$ THz is 23\% higher.
If the phonon DOS $F(\omega)$ function was used to compute $\omega_{\rm ln}$ value, instead of the Eliashberg function, it would be about 40\% higher, 1.45 THz (69.6 K) in the relativistic case. Thus, enhancement of the Eliashberg function over the phonon DOS for lower $\omega$ is responsible for such a low $\omega_{\rm ln}$. 
Strong-coupling correcting factors are $f_1\cdot f_2=1.12$ (with SOC) and $f_1\cdot f_2=1.08$ (without SOC). 
Taking the value of the Coulomb pseudopotential parameter $\mu^* = 0.10$ we obtain the critical temperature of $T_c = 5.43$~K (with SOC), which is in good agreement with the experimental one, equal to $T_c = 5.12$~K. 
The spin-orbit coupling increases $T_c$ in ScAu$_2$Al, as the scalar-relativistic value is 4.57~K.

Although the electron-phonon coupling constant $\lambda = 1.25$ is record high in the family of Heusler compounds, the critical temperature of ScAu$_2$Al is not that high. The reason for this is the remarkably low $\omega_{\rm ln} \simeq 52$~K, only 12\% increased by the strong coupling correction factors $f_1\cdot f_2=1.12$, when $T_c$ is calculated.
This peculiar electron-phonon spectral function, with low frequency enhancement, together with the large value of $\lambda$ is responsible for the significant underestimate of the electron-phonon coupling constant made on the basis of experimental $T_c$, Debye temperature 
($\theta_D = 180$~K) and McMillan formula in Ref.~\cite{Bag2022}.
Due to the highly non-Debye phonon spectrum and strong coupling, the McMillan formula is very inaccurate for ScSu$_2$Al, which explains why much smaller $\lambda = 0.77$ was obtained there.

Our calculations also demonstrate that in the Heusler family of superconductors we reach the strong-coupling regime with $\lambda > 1$ in the compound with the highest $T_c$, like in the other intermetallic families of superconductors.

\subsection{Superconductivity from the SCDFT}\label{sec:sctk}
The motivations for extending our calculations and applying the density functional theory for superconductors method (SCDFT) to ScAu$_2$Al were threefold. First, we avoid using the arbitrary value of the Coulomb pseudopotential $\mu^*$, as in the SCTK package~\cite{Kawamura2017} Coulomb interactions are directly included in the calculations.
This allows us to verify whether the computed strong electron-phonon coupling indeed leads to a critical temperature value of the order of 5~K, with no external parameters. 
Second, SCDFT calculations allow us to investigate the superconducting gap structure in the reciprocal space. 
This is especially interesting as ScAu$_2$Al occurred to be a two-band material, with two large Fermi surface sheets. Therefore, it is possible that different strengths of the electron-phonon interaction on those two sheets will lead to the formation of two different superconducting gaps, as is observed, e.g. in Pb \cite{Ruby2015}.
Two-gap superconductivity was recently suggested in a noncentrosymmetric half-Heusler compound LuPdBi \cite{Ishihara2021} based on upper critical field and penetration depth measurements.

Third, in the experimental studies~\cite{Bag2022} a quadratic temperature dependence of resistivity, $\rho(T) = \rho_0 + AT^2$ was observed. Such a quadratic temperature dependence may be induced by the presence of weak ferromagnetic spin fluctuations (SF), which would additionally compete with superconductivity. The presence of SF may be taken into account in SCDFT calculations within SCTK thus we aimed to verify whether SF have a strong influence on superconductivity in ScAu$_2$Al. All calculations reported in this section include the spin-orbit coupling. 

The superconductivity in SCTK is described with the equation for the superconducting gap function $\Delta_{n\mathbf{k}}$:
\begin{align}\label{eq:delta}
    \Delta_{n\mathbf{k}} = & -\frac{1}{2}\sum_{n'\mathbf{k}'} \frac{K_{n\mathbf{k}n'\mathbf{k}'}(\xi_{n\mathbf{k}},\xi_{n'\mathbf{k}'})}{1+Z_{n\mathbf{k}}(\xi_{n\mathbf{k}})} \nonumber \\
    & \times \frac{\Delta_{n'\mathbf{k}'}}{\sqrt{\xi_{n'\mathbf{k}'}^2+\Delta_{n'\mathbf{k}'}^2}} \tanh \frac{\sqrt{\xi_{n'\mathbf{k}'}^2+\Delta_{n'\mathbf{k}'}^2}}{2T},
\end{align}
where $\xi_{n\mathbf{k}}$ is the Kohn-Sham eigenvalue at $E_F=0$ in the band $n$. $T_c$ is found when the gap vanishes at a given temperature. The integration kernel $K$ 
\begin{align}
    K_{n\mathbf{k}n'\mathbf{k}'}(\xi,\xi') & \equiv K^{ep}_{n\mathbf{k}n'\mathbf{k}'}(\xi,\xi') \nonumber \\
    & + K^{ee}_{n\mathbf{k}n'\mathbf{k}'}(\xi,\xi') + K^{sf}_{n\mathbf{k}n'\mathbf{k}'}(\xi,\xi')\label{eq_K}   
\end{align}
%\begin{align}
%    K_{n\mathbf{k}n'\mathbf{k}'}(\xi,\xi') \equiv K^{ep}_{n\mathbf{k}n'\mathbf{k}'}(\xi,\xi') + %K^{ee}_{n\mathbf{k}n'\mathbf{k}'}(\xi,\xi') \label{eq_K}   
%\end{align}
consist of electron-phonon, electron-electron and spin fluctuations terms. The renormalization factor $Z$:
\begin{equation}
    Z_{n\mathbf{k}}(\xi) = Z^{ep}_{n\mathbf{k}}(\xi) + Z^{sf}_{n\mathbf{k}}(\xi)   
\end{equation}
has only the electron-phonon and SF terms. The screened Coulomb interaction is calculated with the random phase approximation. A more detailed description of the theory can be found in \cite{Kawamura2020,Kawamura2017}.

We first discuss the results of calculations with the spin fluctuation terms neglected.
The main results of the calculations are shown in Table~\ref{tab_sctk}.
The superconducting critical temperature,
calculated from SCDFT based on the gap equation (\ref{eq:delta}) and the condition of vanishing gap at the transition point, is $T_c = 5.16$~K.
This value is almost identical to $T_c = 5.12$~ K, reported in the experiment in Ref. {\it et al.}~\cite{Bag2022}, and larger than 
$T_c = 4.4$~K previously reported for that system \cite{Poole2000,Winiarski2021}.
This confirms that ScAu$_2$Al has intrinsically the highest critical temperature among the Heusler compounds studied so far, and the lower value, reported in earlier studies~\cite{Poole2000}, may be due to disorder in the samples. 

Figure~\ref{fig_sctk} presents the $\mathbf{k}$-space distribution of the key parameters for the superconductivity of ScAu$_2$Al: the value of the electron-phonon coupling parameter $\lambda_{n\mathbf{k}}$ in panel (a), the screened Coulomb repulsion parameter $\mu_{n\mathbf{k}}$ in panel (b), and the wave-vector dependence of the superconducting gap $\Delta_{n\mathbf{k}}$ in panel (c), each plotted on the Fermi surface of the compound.
Panels (d), (e) and (f) show those quantities in a form of histograms. 

First of all, superconductivity is fully gaped, there are no points or lines with nodes.
A moderate level of anisotropy and a quite large spread of the discussed quantities is observed, more pronounced for the Fermi surface associated with the first band.
The electron-phonon coupling parameter $\lambda_{n\mathbf{k}}$ takes values between 0.7 and 2.0, and the regions with stronger coupling also have weaker screened electron repulsion parameter $\mu_{n\mathbf{k}}$. This results in a distribution of $\Delta_{n\mathbf{k}}$ with quite a noticeable spread, the ratio of the highest to the lowest values reaches
about 2.
Most of the FS regions exhibit $4 < 2\Delta/k_BT_c < 5$, much above the BCS weak-coupling limit of 3.53.
The regions with the strongest electron-phonon coupling, which includes the nested parts of the FS according to Fig.~\ref{fig_susc}, correspond to the largest $\Delta_{n\mathbf{k}}$.
For example, the gap in the $\Gamma$-K direction, where the acoustic phonon modes are strongly coupled and the nesting function is enhanced, has a higher value than in $\Gamma$-L. 
The gap near the L point has moderate values, as well as relatively small phonon linewidths seen in Fig.\ref{fig_gamlines}, which shows that the flat band at L does not affect the superconductivity much. 

\begin{table}[t]
	\caption{Results of SCDFT calculations in a form of average parameters for FS sheets corresponding to band 1, band 2, and the global average, obtained with and without spin fluctuations (all with spin-orbit coupling): $\mu$ -  Coulomb repulsion parameter, $\Delta$ - superconducting gap, $\lambda$ - electron-phonon coupling constant. $T_c$ is the superconducting critical temperature.\label{tab_sctk}}
\begin{center}
\begin{ruledtabular}
\begin{tabular}{lcc}
      & w/o SF & with SF\\
      \hline
    $\overline{\mu}_{b_1}$ & 0.270 & 0.297\\
    $\overline{\mu}_{b_2}$ & 0.273 & 0.298\\
    $\overline{\mu}$ & 0.271 & 0.297\\
    $\overline{\Delta}_{b_1}$ (meV) & 0.909 & 0.840\\
    $\overline{\Delta}_{b_2}$ (meV) & 0.947 & 0.878\\
    $\overline{\Delta}$ (meV) & 0.921 & 0.854\\
    $2\overline{\Delta}_{b_1}/k_BT_c$ & 4.093 & 4.075\\
    $2\overline{\Delta}_{b_2}/k_BT_c$ & 4.273 & 4.264\\
    $2\overline{\Delta}/k_BT_c$ & 4.138 & 4.119\\
    $\overline{\lambda}_{b_1}$ & 1.355 & 1.355\\
    $\overline{\lambda}_{b_2}$ & 1.387 & 1.387\\
    $\overline{\lambda}$ & 1.364 & 1.364\\
    $T_c$ (K) & 5.16 & 4.79\\
\end{tabular}
\end{ruledtabular}
\end{center}
\end{table}

\begin{figure*}[t]
	\centering
	\includegraphics[width=0.75\textwidth]{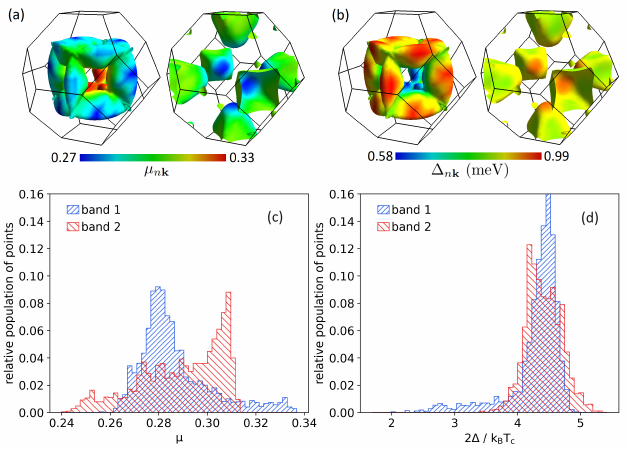}
	\caption{The $n\mathbf{k}$ dependent (a) Coulomb potential and (b) superconducting gap function at $T=0$~K. Histograms in (c) and (d) present the distribution of the quantities that were plotted on the Fermi surface. These results were computed with SOC and SF.\label{fig_sctk_sf}}	
\end{figure*}

From the k-dependent values we calculate the average superconducting gaps for each band, by weighting $\Delta_{n\mathbf{k}}$ by the k-resolved density of states at $E_F$
% \begin{equation}
%     \Delta_n = \frac{\sum_{\mathbf{k}} \Delta_{n\mathbf{k}}N^{mv}_{\mathbf{k}}}{\sum_{\mathbf{k}} N^{mv}_{\mathbf{k}}}.
% \end{equation}
\begin{equation}
   \overline{\Delta}_n = \frac{\sum_{\mathbf{k}} \Delta_{n\mathbf{k}}\delta(E_{\mathbf{k}} - E_F)}{\sum_{\mathbf{k}}{\delta(E_{\mathbf{k}} - E_F})}.
\end{equation}
The $\delta(E_{\mathbf{k}} - E_F)$ functions are computed using the Marzari-Vanderbilt cold smearing method \cite{Marzari1999,marzari2023}, more details are given in the Supplemental Material~\cite{suppl}. 
The average Coulomb repulsion parameters $\overline{\mu}_n$ and electron-phonon coupling constants $\overline{\lambda}_n$ were computed in the same way, and $n = \{b_1, b_2\}$ stand for the Fermi surface sheet of band 1 and band 2. 
The global averages are also computed by weighting each FS sheet contribution.
The results are collected in Table~\ref{tab_sctk}.
The average values of the superconducting gaps confirm the strong-coupling character of superconductivity in ScAu$_2$Al, as 
$2\overline{\Delta}_{b_1}/k_BT_c = 4.09$ 
and
$2\overline{\Delta}_{b_2}/k_BT_c = 4.27$, both considerably above the BCS value.
As the first FS sheet contributes about 75\% to $N(E_F)$ due to lower Fermi velocities (see Fig.~\ref{fig_bandsfs}), the global average is closer to the 
$\Delta_{b_1}$, with $2\overline{\Delta}/k_BT_c = 4.14$.
The Fermi surface average electron-phonon coupling parameters $\overline{\lambda}_{n}$ are similar for both FS sheets, and the global average $\overline{\lambda} = 1.36$.
This value is 9\% larger than the value computed from the Eliashberg function in the previous section ($\lambda = 1.25$). The difference is the result of different numerical procedures of calculations of $\lambda$ for the two cases, including unshifted {\it vs.} shifted $\mathbf{q}$-point meshes or smearing {\it vs.} tetrahedron integration methods.
Nevertheless, the difference is not significant and even stronger electron-phonon coupling is predicted by the SCDFT calculations.

The screened Coulomb interaction parameter $\mu_{n\mathbf{k}}$ is shown in Fig.~\ref{fig_sctk}(b). It is obtained from the electron-electron part of the kernel $K^{ee}_{n\mathbf{k}n'\mathbf{k}'}$ [see eq. (\ref{eq_K})] by summing over $\{n'\mathbf{k}'\}$ on the Fermi surface. 
By integrating it again, one obtains the average Coulomb repulsion parameter $\overline{\mu}$, shown in Table~\ref{tab_sctk}, which is a dimensionless product of the average interaction kernel times the density of states at $E_F$~\cite{Kawamura2020}.
Both FS sheets have similar values, and the overall average is
$\overline{\mu} = 0.27$.
Compared to elemental metals that build our structure,
Au with the filled d-shell well below the Fermi energy has a lower $\overline{\mu}=0.14$, metallic Al has a similar $\overline{\mu}= 0.25$
and Sc with an open 3d shell has a much higher $\overline{\mu}=0.52$ \cite{Kawamura2020}.

\begin{figure}[b]
	\centering
	\includegraphics[width=0.99\columnwidth]{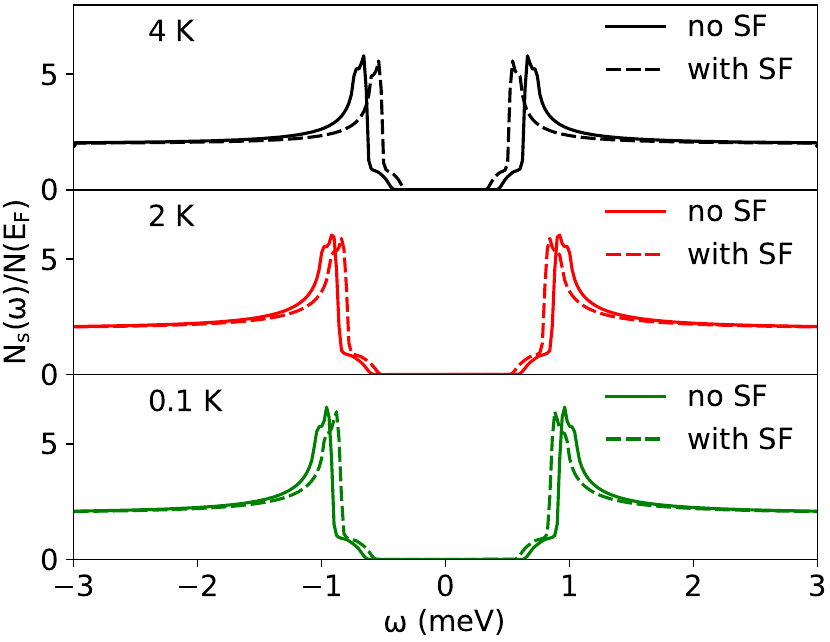}
	\caption{Quasiparticle spectrum as a function of temperature. Its shape is visibly affected by the anisotropy of the superconducting gap.}\label{fig_quasi}	
\end{figure}

In the superconducting state, the effect of Coulomb repulsive interactions is weakened by retardation effects due to different energy scales of electrons and phonons, and this effect is naturally included in the SCDFT calculations~\cite{Kawamura2020}. 
In the isotropic 
Eliashberg theory or in the Allen-Dynes formula 
retardation is approximately described using the $\mu^*$ parameter~\cite{Bogoljubov1958,MorelAnderson1962,carbotte1990,chang_nb,gunnarsson2013}
\begin{equation}\label{eq:mu*}
    \mu^* = \frac{\mu}{1+\mu\ \ln(E_{\rm el}/\omega_{\rm ph})}.
\end{equation}
The $E_{\rm el}$ and $\omega_{\rm ph}$ are usually taken as the electronic and phononic cutoff energies, of the order of Fermi energy or bandwidth for the former and $\omega_{\rm max}$ or $\theta_D$ for the latter (due to the logarithm in the denominator the slightly different definitions do not affect $\mu^*$ much).
Based on the calculated $\overline{\mu}$ we may estimate $\mu^*$. 
With an electronic bandwidth of 10 eV and a maximum phonon frequency of 7.5 THz (31 meV) one obtains $\mu^* = 0.105$. 
If the Debye temperature of 180~K is used, instead of $\omega_{\rm max}$, a smaller $\mu^* = 0.092$ is obtained, but all close to the standard approximation of $\mu^* = 0.10$.

Now, let us proceed to the analysis of the spin fluctuation effect. 
Figure~\ref{fig_sctk_sf}(a) shows the {\bf k}-space distribution of the repulsion parameter $\mu_{n\mathbf{k}}$, which now includes both the Coulomb and spin-fluctuation kernels. In panel (b) the superconducting gap is presented, and panels (c) and (d) show their histograms. SF term does not influence $\lambda_{n\mathbf{k}}$, so it is not repeated after Fig.~\ref{fig_sctk}.
The critical temperature drops a little, to $T_c = 4.79$~K, now slightly lower than the experimental result of 5.12~K in Bag {\it et al.}~\cite{Bag2022}, yet still larger than 4.4~K in an earlier report \cite{Poole2000}. 
The repulsion parameter  $\mu_{n\mathbf{k}}$ increases, its {\bf k}-dependence becomes a bimodal distribution, with peaks in the histograms shifted between the two bands. Nevertheless the average values for the two FS sheets become equal (see Table~\ref{tab_sctk}) and increase in about 10\% due to SF to $\overline{\mu} = 0.30$\footnote{If we would like to compare the effective $\mu^*$ parameter including the retardation effects, eq.(\ref{eq:mu*}), the increase of $\mu^*$ due to SF is  about 5\%}. 
In elemental metallic scandium, where strong spin fluctuations are present, the SF contribution to the repulsion parameter is much larger, $\overline{\mu}_{{SF}} = 0.976$ \cite{Kawamura2020}, which makes the compound non-superconducting.
The small increase in $\overline{\mu}$ in ScAu$_2$Al is responsible for the small decrease in $T_c$ and in the superconducting gaps, compared to the no-SF calculations; however, the ratios $2\overline{\Delta}_{{n}}/k_BT_c$ remain practically unchanged. 
From our calculations, we may conclude that the spin fluctuations are weak in ScAu$_2$Al.

\begin{figure}[t!]
	\centering
	\includegraphics[width=0.99\columnwidth]{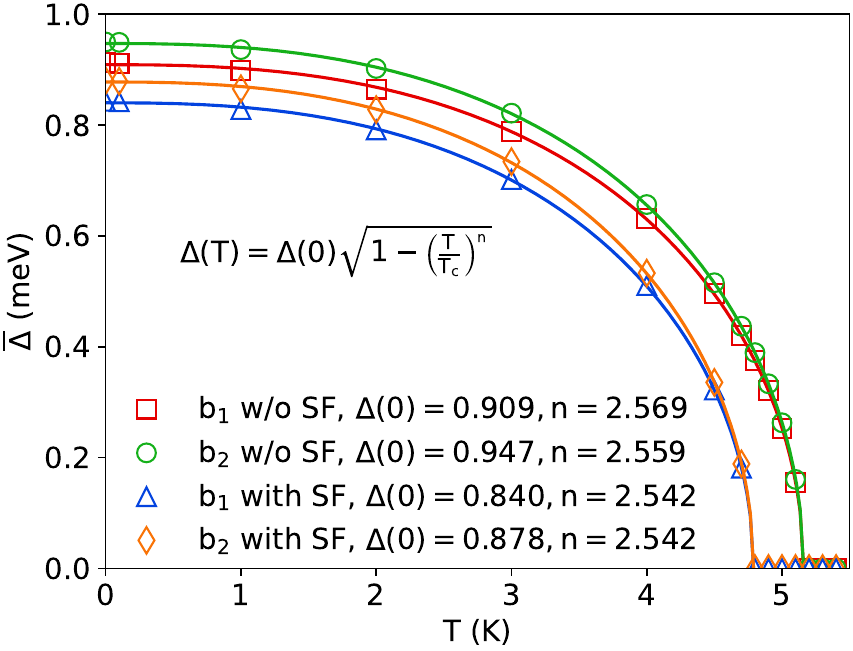}
	\caption{Temperature dependence of the average values of superconducting gaps, calculated without and with spin fluctuations. Lines are fits to the formula (\ref{eq:gap}).}\label{fig_gap}	
\end{figure}

In the final step of our analysis of the superconducting properties of ScAu$_2$Al, we have calculated the temperature-dependent quasiparticle density of states in a superconducting phase (see Ref.~\cite{Kawamura2017} for details) and the temperature evolution of the superconducting gap. Figure~\ref{fig_quasi} shows the quasiparticle DOS spectrum for 0.1~K, 2~K and 4~K, computed with and without spin fluctiations (both include SOC). As one can clearly see, it deviates from the BCS-like singularity at the energy gap and it is a consequence of the {\bf k-}dependence and energy spread of the superconducting gaps, discussed above. The small hump at its beginning comes from the ''tail'' seen on $\Delta_{n\mathbf{k}}$ histograms in Figs.~\ref{fig_sctk} and \ref{fig_sctk_sf}.
Tunneling conductance measurements should be performed to verify our calculations and further investigate the gap anisotropy.
Figure~\ref{fig_gap} presents the temperature dependence of the average value of superconducting gaps for bands 1 and 2, computed with and without SF (both include SOC).
The gaps were fitted to the equation:
\begin{equation}
\label{eq:gap}
\Delta\left(T\right)=\Delta\left(0\right)\sqrt{1-\left(\frac{T}{T_c}\right)^{n}}, 
\end{equation}
where in the BCS model $n \simeq 3.0$~\cite{kuderowicz2021}.
Here, the fitting gives a smaller exponent of $n \simeq 2.5$ with SF and $n \simeq 2.6$ without SF.

\section{Summary}
To summarize, we have presented calculations of the electronic structure, phonons, electron-phonon coupling functions, superconducting critical temperature, and superconducting gap in ScAu$_2$Al. 
Spin-orbit coupling is found to have an important effect on the electronic structure near the Fermi level, because it lifts the degeneracy of the bands, removing the van Hove singularity from the vicinity of the Fermi level. 
Sc-d states contribute approximately 50\% of the density of states at the Fermi level, while the Al-p states contribute 30\% with Au-d and Au-s constituting the rest. 
Due to a large difference of the Au atom mass compared to Sc and Al, phonons form two groups of modes, separated by a large gap of approximately 2 THz. 
Although Sc is heavier than Al, phonon modes dominated by its vibrations have higher frequencies due to the stronger bonding of Sc in the structure.
The most important feature of the phonon structure to provide the strong electron-phonon coupling in this compound is a low-dispersive acoustic branch in the $\Gamma$-X direction that has significantly lower frequencies than the rest of the spectrum and is dominated by Au vibrations.
The SOC further softens phonons in that branch, resulting in a large value of the electron-phonon coupling constant $\lambda=1.25$. 
Therefore, ScAu$_2$Al can be classified as a strong coupling superconductor. 
Using the Allen-Dynes formula with strong coupling correction factors, we obtained $T_c=5.43$ K when using a typical value of $\mu^*=0.10$.

These conclusions are then strengthened by the SCDFT calculations, in which the depairing effects of Coulomb interactions and spin fluctuations are taken into account.
The average electron-phonon coupling parameter, obtained for SCDFT calculations, confirmed that the coupling is strong, giving a slightly larger $\lambda=1.36$. 
If SF are neglected, the critical temperature of $T_c=5.16$ K is obtained, and the inclusion of SF slightly decreases the critical temperature to $T_c=4.79$ K.
The obtained $T_c$ is close to the experimental one of 5.12~K, confirming the electron-phonon coupling mechanism of superconductivity in ScAu$_2$Al and showing that spin fluctuations are weak.
Analysis of the Fermi surface and superconducting gaps show that ScAu$_2$Al is a two-band superconductor with nodeless gaps and a moderate level of gap anisotropy. The values of $2\overline{\Delta}/k_BT_c$ are 4.09 and 4.27 for the two Fermi surface sheets, above the weak coupling limit of 3.53.
The calculated quasiparticle density of states is strongly affected by the large spread of superconducting energy gap values and its anisotropy, opening a possibility of its experimental validation through the tunneling spectroscopy measurements.

\section*{Acknowledgements}
This work was supported by the National Science Centre (Poland), project no. 2017/26/E/ST3/00119.
We gratefully acknowledge Polish high-performance computing infrastructure PLGrid (HPC Centers: ACK Cyfronet AGH) for providing computer facilities and support within computational grants no. PLG/2022/015620 and PLG/2023/016451.
We thank Kamil Kutorasiński for help in evaluating the nesting function.

\bibliography{refs}

\newpage

\onecolumngrid

\section*{Supplemental Material}

\renewcommand{\thefigure}{{S\arabic{figure}}}
\setcounter{figure} 0

\vspace*{24pt}
\noindent
Supplemental Material contains:\\ \\
% Figure \ref{fig_BZ} of the Brillouin zone with the high symmetry points;\\ 
Figure \ref{fig:w2k} which compares the electronic structure obtained in pseudopotential and all-electron calculations;\\
Figure \ref{fig_phconv} of the phonon dispersion relations and phonon densities of states obtained on 6x6x6 and 8x8x8 q-point grids with additional discussion of the convergence tests for calculations of $\lambda$; \\
Discussion of the specific heat with Fig. \ref{fig_gammafits} of the Sommerfeld coefficient and Debye temperature obtained by fitting the experimental specific heat data as a function of the temperature range, as well as Figs.~\ref{fig_phheat}-\ref{fig_debye} that demonstrate the inapplicability of the Debye model for a description of the lattice specific heat above 2~K; \\
Additional details of the cold smearing method used to calculate the average values of the quantities obtained in the SCDFT calculations;\\
Description of the video files with the animations of atomic displacements in the selected phonon modes.

\subsection{Verification of pseudopotentials}

\begin{figure}[H]
	\centering
	\includegraphics[width=0.45\columnwidth]{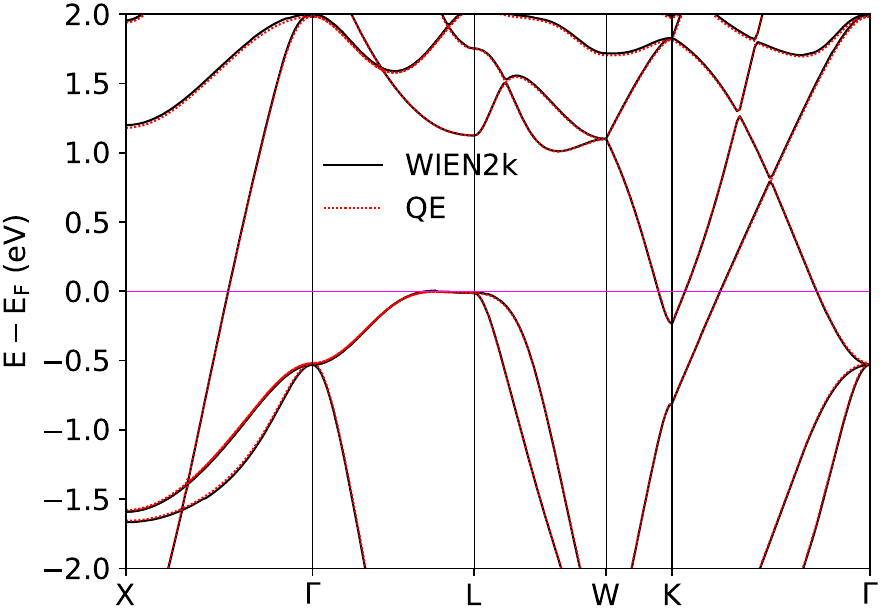}\\
 
    \includegraphics[width=0.45\columnwidth]{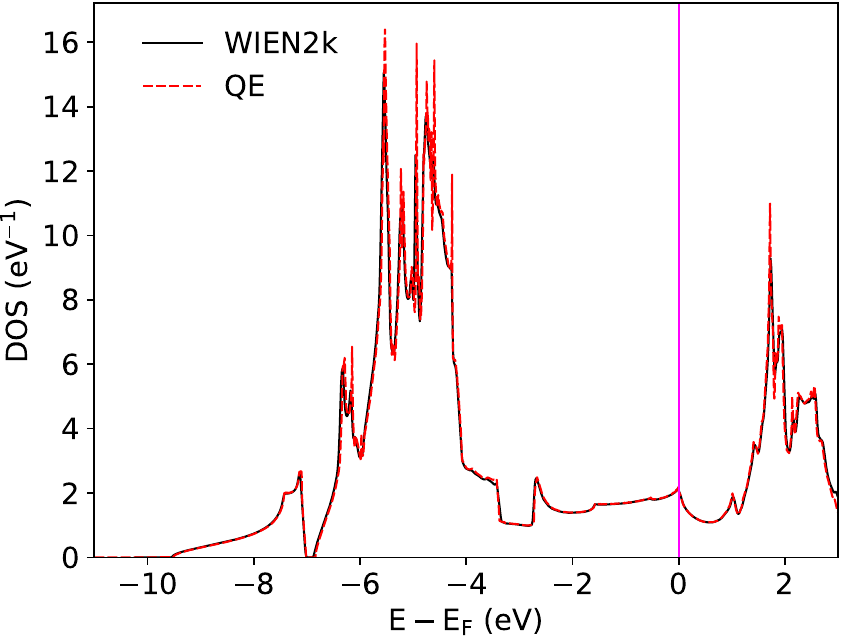}
	\caption{Comparison of the (scalar-relativistic) density of states and electronic dispersion relations in ScAu$_2$Al obtained from pseudopotential calculations in {\sc quantum espresso} (QE) and all-electron calculations in {\sc wien2k}. \label{fig:w2k}}
\end{figure}

To verify that the chosen pseudopotentials and cutoff energies provide an accurate description of the electronic structure, the computed electronic dispersion relations and densities of states were compared to the results of all-electron full-potential linearized augmented plane wave calculations, performed using the {\sc wien2k} package. 
Figure~\ref{fig:w2k} shows that the obtained results are identical, validating the choice of pseudopotentials and other computational parameters. 

\newpage
\subsection{Convergence tests for phonon and $\lambda$ calculations}

The convergence of calculations of the phonon structure and electron-phonon coupling parameter $\lambda$ was verified in three steps.
\begin{enumerate}

\item QE calculations of Eliashberg functions were done for an unshifted 6x6x6 {\bf q}-point mesh (16 inequivalent points for which dynamical matrices are calculated), based on the self-consistent cycle on the 12x12x12 {\bf k}-point mesh. Integrations were carried out using the cold smearing method, with the smearing parameter $\sigma=0.02$ Ry. 
The eigenvalues for the electron-phonon interaction matrix elements were calculated on a 24x24x24 {\bf k}-point mesh. 
The Eliashberg functions were then calculated with $\sigma=0.01$ Ry. 
In the scalar-relativistic case it resulted in $\lambda$ = 0.974, whereas with SOC the calculated $\lambda$ = 1.25. This set of mesh parameters is labeled "q6".
The convergence of the {\bf k} and {\bf q}-point meshes and convergence against the smearing parameter was then checked by repeating the entire calculation cycle on a denser meshes: 8x8x8 q-point + 16x16x16 {\bf k}-point for the self-consistent cycle and 32x32x32 {\bf k}-point for eigenvalues for the electron-phonon matrix elements (“q8” in short), with the same smearing parameters (0.02 and 0.01 Ry). The comparison of phonon DOS and dispersion relations is shown in Fig.~\ref{fig_phconv}.

\begin{figure}[H]
	\centering
	\includegraphics[width=0.45\columnwidth]{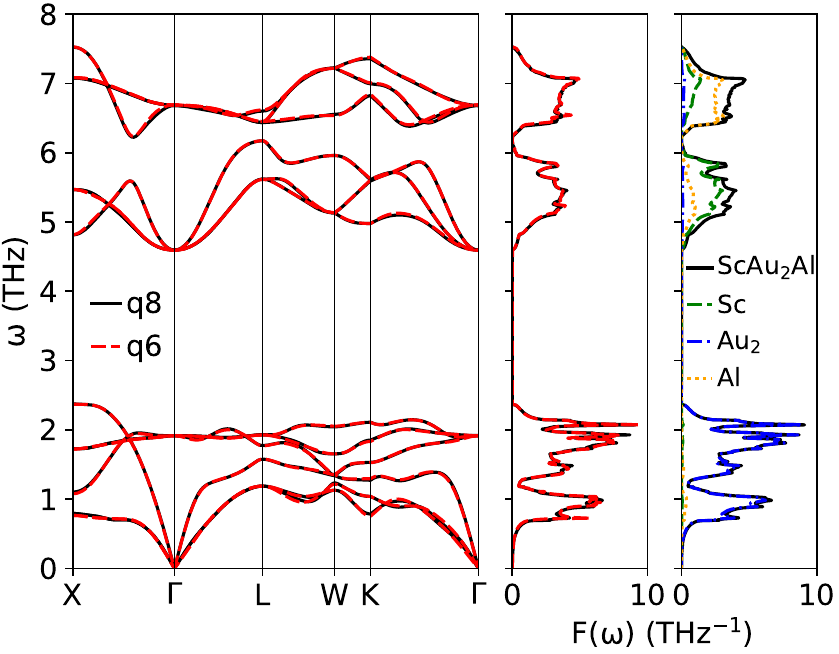}
	\caption{Phonon $\mathbf{q}$ grid convergence test - comparison of the phonon dispersion relations and densities of states obtained for $6^3$ and $8^3$ grids. \label{fig_phconv}}
\end{figure}

As can be seen, the results are almost exactly the same. The “q6” set resulted in $\lambda$ = 0.974, whereas the “q8” in $\lambda$ = 0.991, so the difference (1.7\%) is negligible, confirming that the "q6" mesh parameters and selected smearing give convergent results.
Calculations including SOC were done only for the “q6” mesh.

\item In the next step of convergence tests, we recalculated the relativistic Eliashberg function of the “q6” mesh using a twice larger smearing parameter of 0.02 Ry (in the last step of calculations, where the Eliashberg function is calculated based on the electron-phonon interaction matrix elements). 
The larger smearing resulted in 4\% lower $\lambda$ = 1.20.
During these calculations the total DOS($E_F$) is also calculated and the obtained value was considerably lower (1.7 $eV^{-1}$) from the tetrahedron result (2.0 $eV^{-1}$). As for the smaller smearing parameter of 0.01 Ry the DOS($E_F$) (1.9 $eV^{-1}$) was much closer to the tetrahedron result, the $\lambda$ value obtained for a smaller smearing was considered more accurate and used in our work in further analysis.

\item Finally, an independent set of calculations was performed in order to analyze superconductivity using the SCDFT method (SCTK code). Here, a shifted 6x6x6 {\bf q}-point mesh (28 inequivalent {\bf q} points) + 12x12x12 k-point mesh for self-consistent cycle and 24x24x24 {\bf k}-point mesh for eigenvalues to the electron-phonon coupling calculations were used to obtain the dynamical matrices and the electron-phonon coupling coefficients. In all calculations here the optimized tetrahedron integration method was used. 
Due to different meshes and different integration methods, the obtained Fermi-surface average $\lambda$ = 1.36 differs from $\lambda$ = 1.25 in the QE calculations on the unshifted mesh with cold-smearing integrations. However, the difference (9\%) is not large; obtained value confirms the strong electron-phonon-coupling regime, and the 9\% difference between the two methods may be considered as a measure of the numerical accuracy of our calculations.

% Electron-phonon calculations, reported in our work, were done on a 6x6x6 $\mathbf{q}$-pont grid. To ensure that this is a sufficient sampling, we have performed scalar-relativistic calculations of the phonon structure on a denser, 8x8x8 grid. The phonon dispersions and densities of states are compared in Fig.~\ref{fig_phconv} and confirm that 6x6x6 sampling is sufficient. 

\end{enumerate}
\newpage
\subsection{Analysis of the specific heat}
\begin{figure}[H]
	\centering
	\includegraphics[width=0.45\columnwidth]{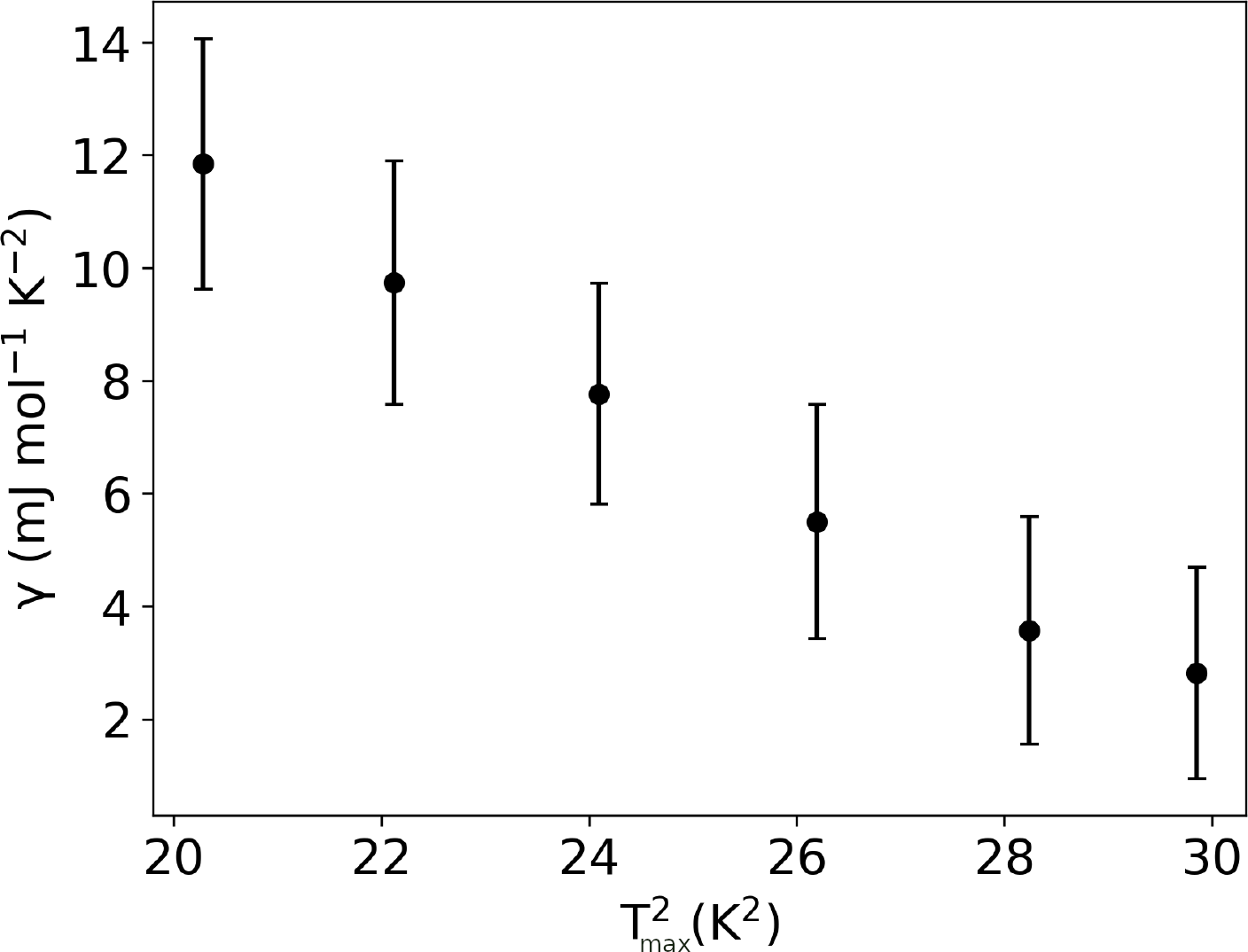}
 \includegraphics[width=0.45\columnwidth]{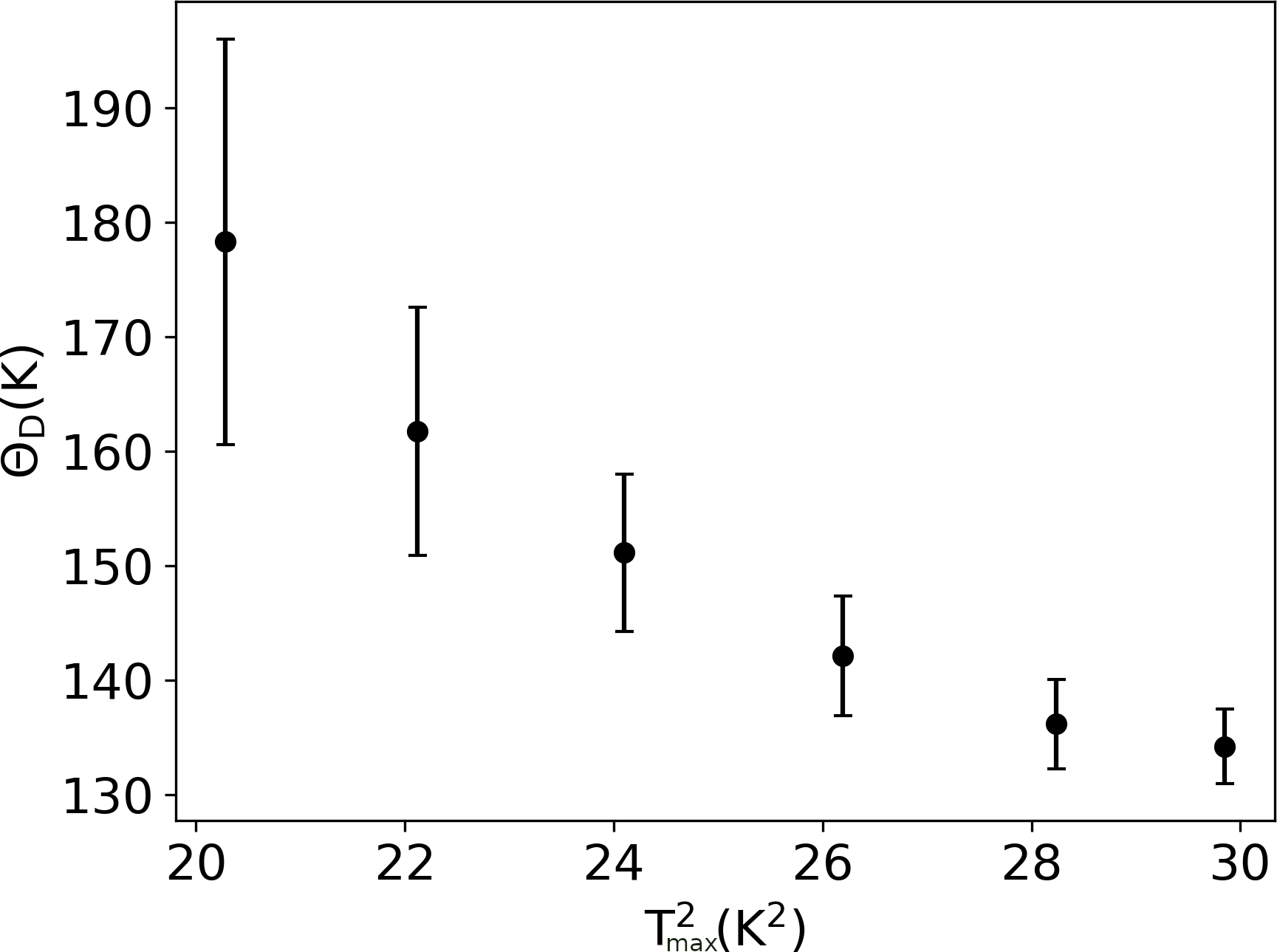}
	\caption{Sommerfeld coefficient $\gamma$ obtained from fits $C/T=\gamma+\beta T^2+\delta T^4$ to the experimental data [37] in the temperature range $(2, T_{\rm max})$, as function of $T_{\rm max}$. Error bars are $\pm$ standard deviation.\label{fig_gammafits}}
\end{figure}

We reanalyzed the experimental heat capacity at constant pressure C$_p$ measured in a magnetic field of 10 kOe and presented in Fig. S4 of the supplemental material to Bag et al., J. Phys.: Condens. Matter 34 (2022) 195403 (Ref. [37] in the main text). 
For the extracted data, we fitted the specific heat with the formula $C/T=\gamma+\beta T^2+\delta T^4$, and the Sommerfeld coefficient $\gamma$ and the Debye temperature $\theta_D$ were obtained ($\theta_D = \left(\frac{12\pi^4 n R}{5\beta}\right)^{1/3}$).
Fits were done in the temperature range 
2.0 K  - $T_{\rm max}$, with 4.5 K $<T_{\rm max}< 5.5$ K ($T_{\rm max}^2$ from 20 K$^2$ to 30 K$^2$). 
Figure~\ref{fig_gammafits} shows the fitted $\gamma$ and $\theta_D$ as a function of $T_{\rm max}^2$. First, the error bars are very large and second,
it appeared that the obtained parameters are strongly sensitive to the fitting range. Sommerfeld parameter varies from 12 mJ/(mol K$^2$) for the narrower fit range to 2.8 mJ/(mol K$^2$) for the broader range, while $\theta_D$ changes from 180~K to 135~K.
The precision of the measurement (authors in [37] mentioned problems with stabilization of the temperature) could have an influence on that; however, more importantly, our phonon calculations showed that the specific heat in ScAu$_2$Al even at low temperatures cannot be described by the Debye model and a power-law formula.

\begin{figure}[H]
	\centering
	\includegraphics[width=0.49\columnwidth]{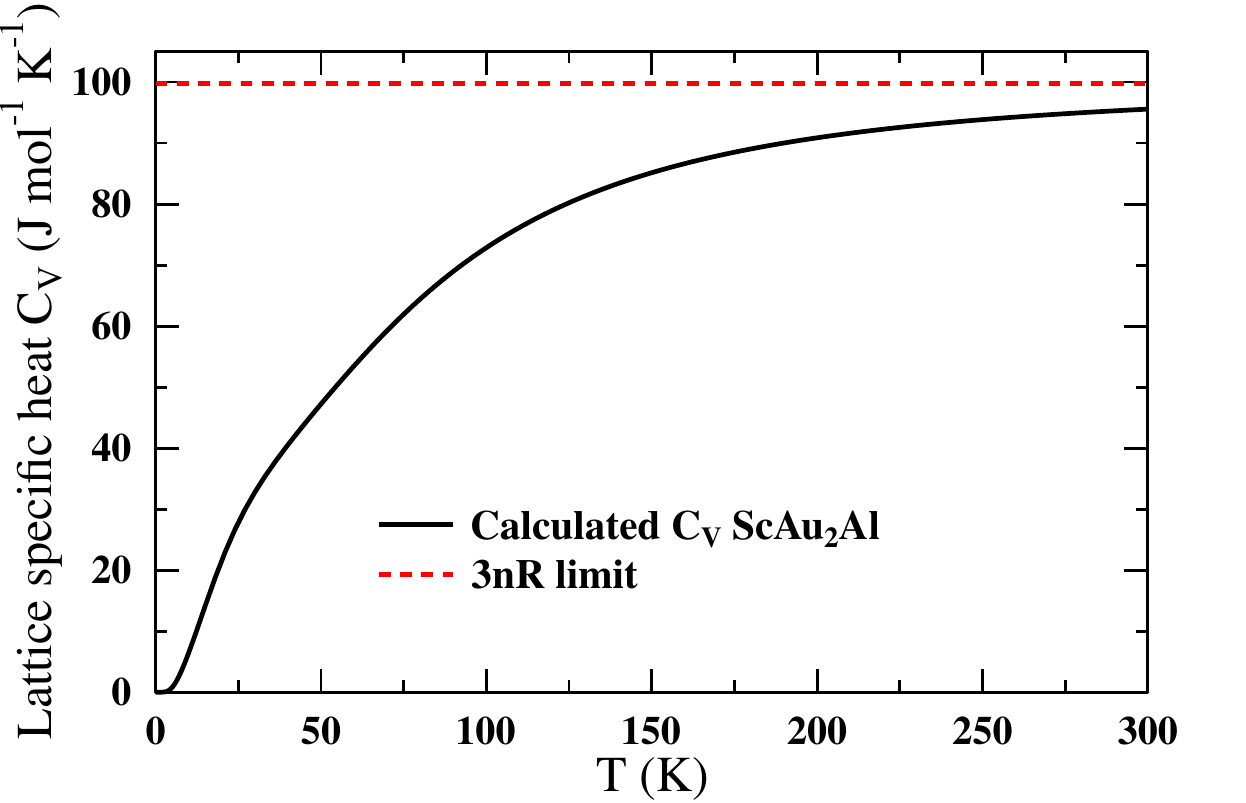}
 	\includegraphics[width=0.49\columnwidth]{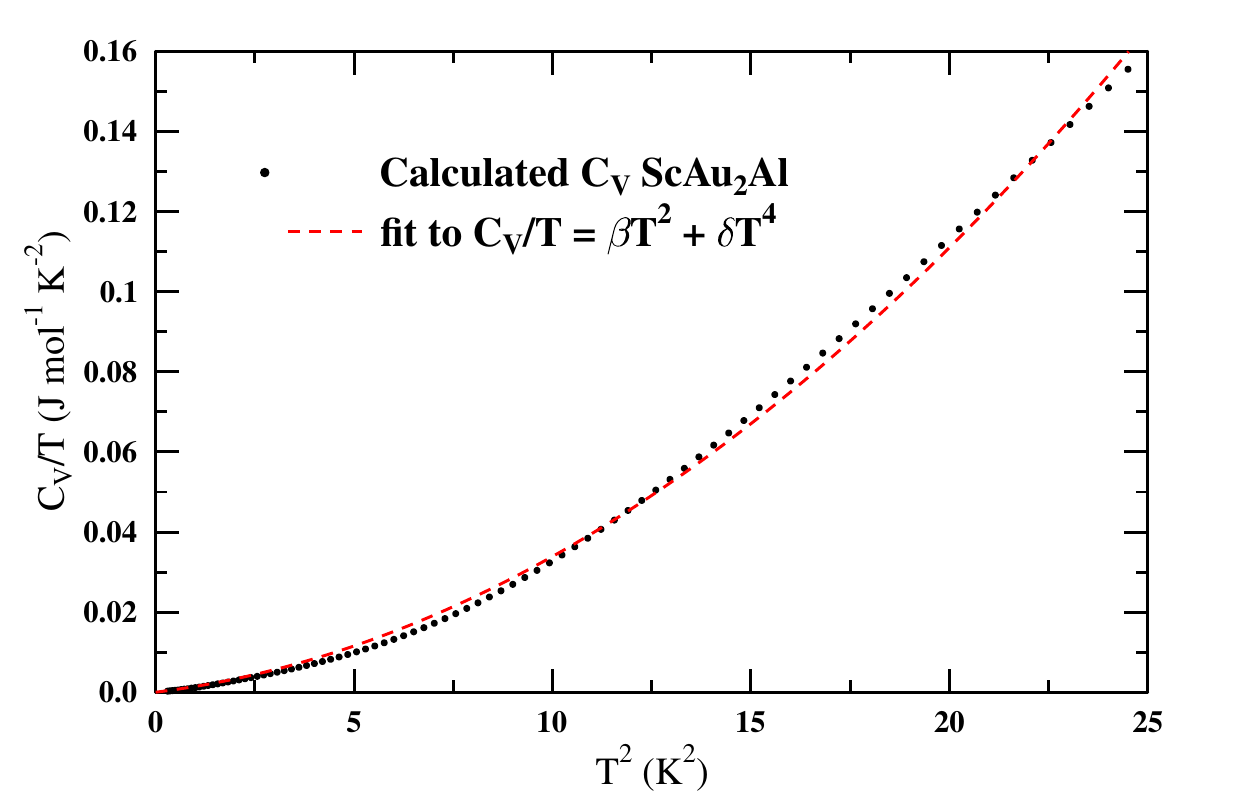}
	\caption{Theoretical constant volume lattice specific heat $C_V$ in ScAu$_2$Al, calculated with Eq.(~\ref{eq:heat}). Left: shown in the broad temperature range; Right: low-temperature behavior of $C_V/T$ {\it versus} $T^2$: points are calculated values, dashed line is a fit to $C_V/T = \beta T^2 + \delta T^4$ formula. In 0 - 5 K range the fitting gives $\beta = 1.226$ (mJ mol$^{-1}$ K$^{-4}$), $\delta = 0.216$ (mJ mol$^{-1}$ K$^{-6}$).\label{fig_phheat}}
\end{figure}

\begin{figure}[H]
	\centering
 \includegraphics[width=0.49\textwidth]{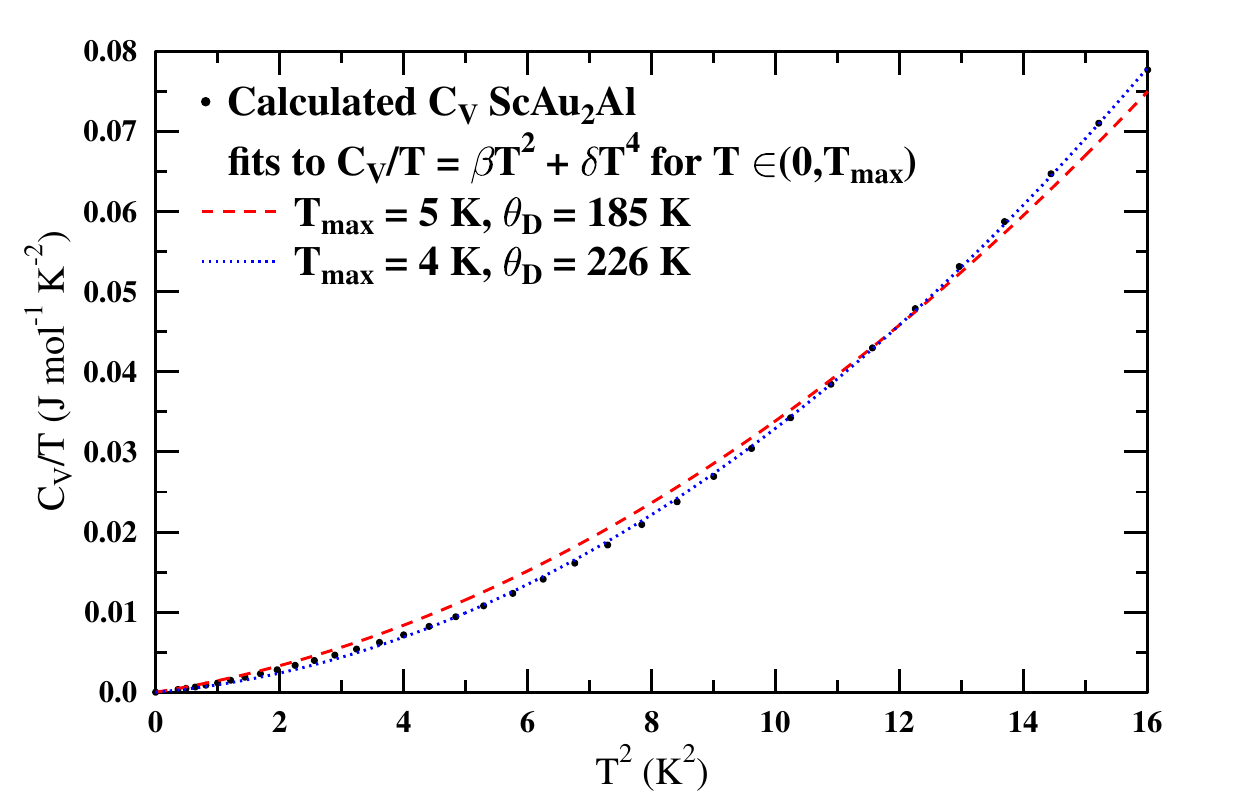}
 \includegraphics[width=0.49\textwidth]{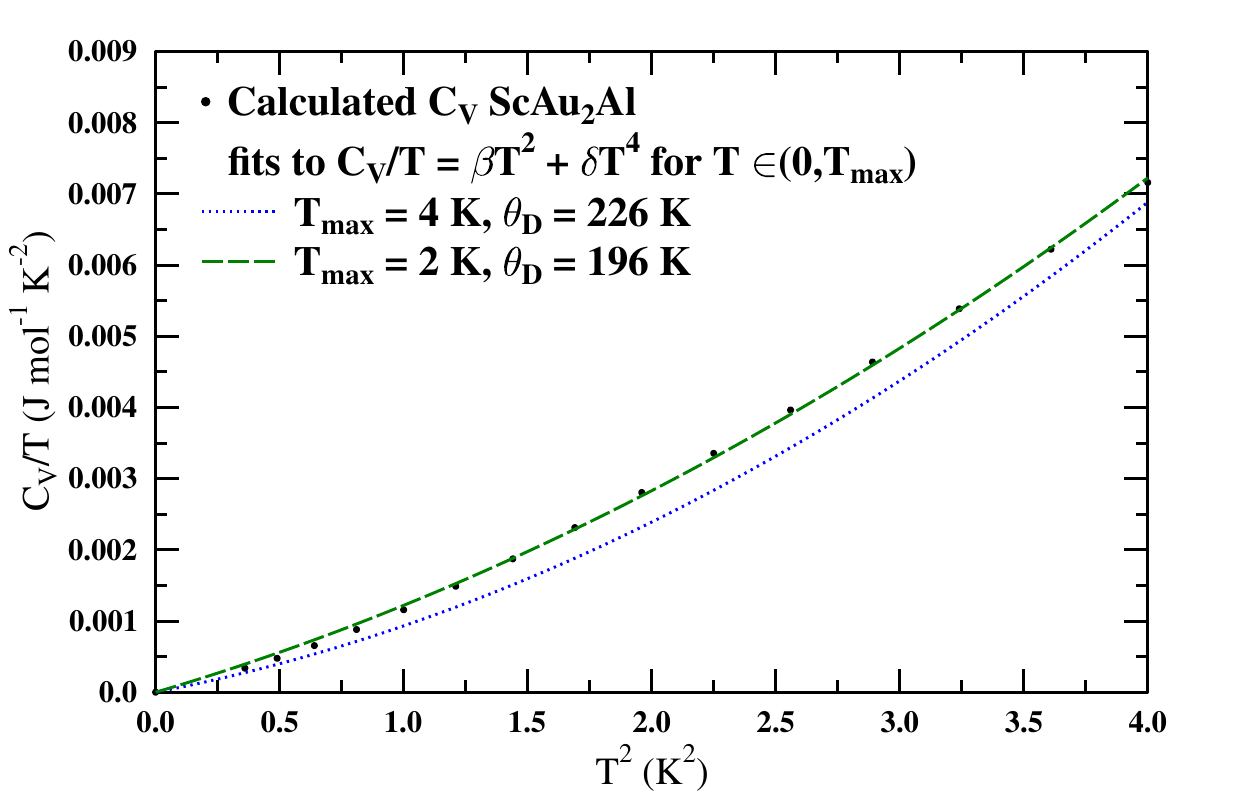}
	\caption{Theoretical constant volume lattice specific heat $C_V$ in ScAu$_2$Al, calculated with Eq.(~\ref{eq:heat}) (points). Lines are fits to the calculated specific heat with the $C_V/T = \beta T^2 + \delta T^4$ formulas in the different temperature ranges $(0, T_{\rm max})$, resulting in different $\theta_D$. Parameters obtained in a larger temperature range fail to describe $C_V$ at lower temperatures. $T_{\rm max} = 5$ and 4~K (left panel), 4 and 2~K (right panel). \label{fig_heat3}}
\end{figure}

\begin{figure}[H]
	\centering
\includegraphics[width=0.49\columnwidth]{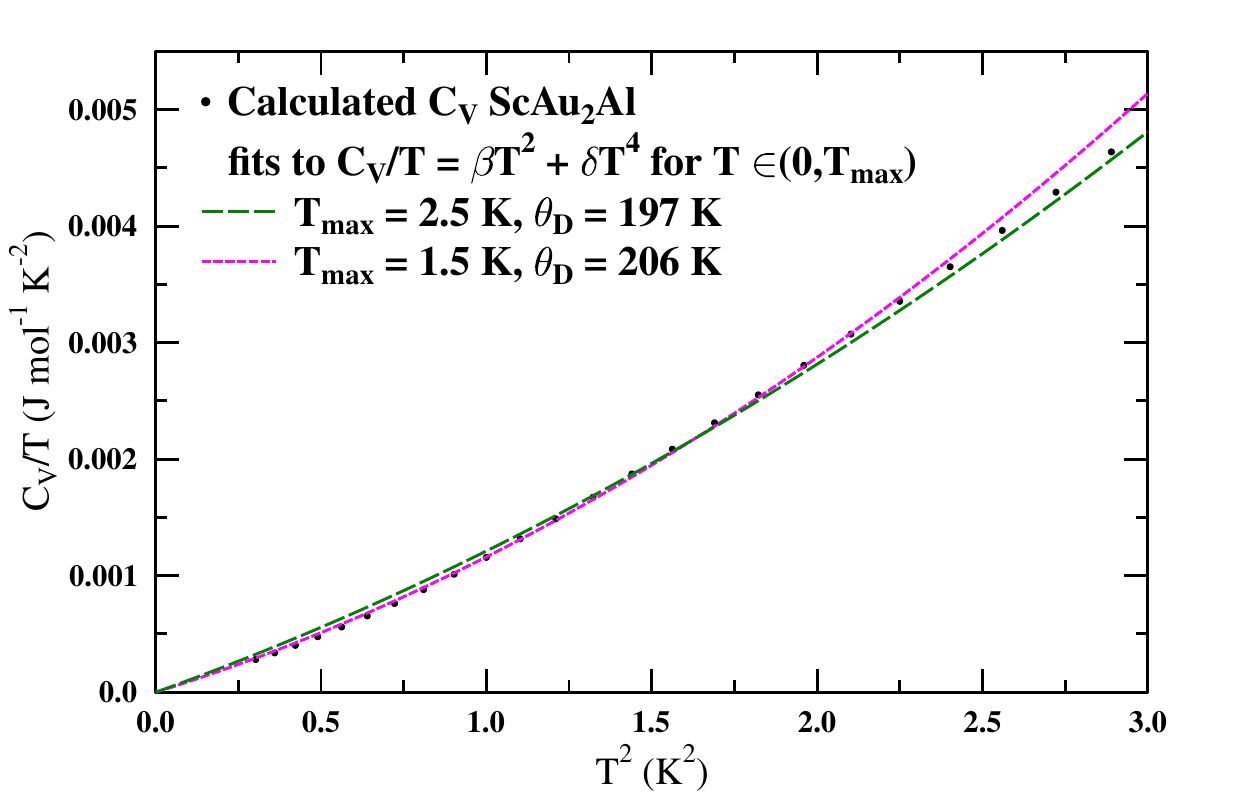}	
 \includegraphics[width=0.49\columnwidth]{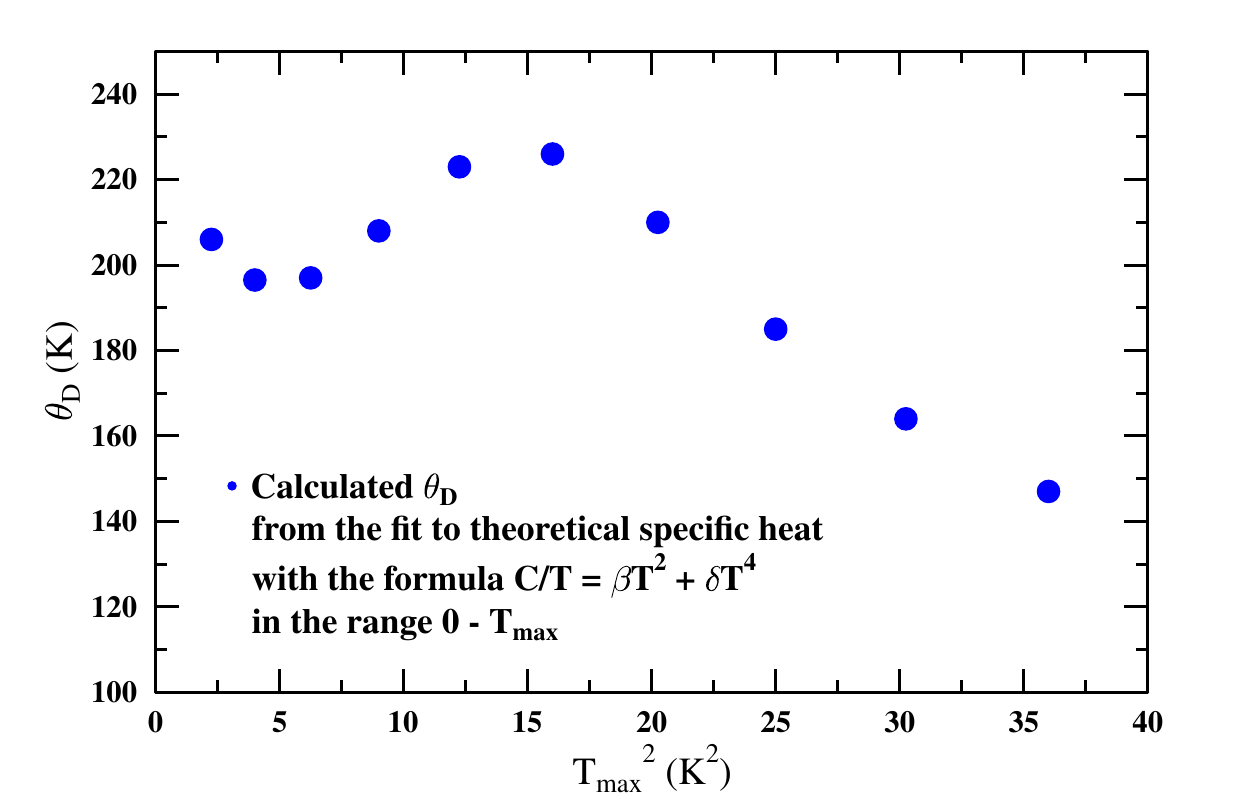}
	\caption{Left panel: same as Fig.~\ref{fig_heat3} with $T_{\rm max} = 2.5$ and 1.5~K. Right panel: $\theta_D$ obtained from fitting the power-law formula $C_V/T = \beta T^2 + \delta T^4$ to the theoretical $C_V$ [Eq.(~\ref{eq:heat})] in the temperature range $(0, T_{\rm max}$ as a function of  $T_{\rm max}$. \label{fig_debye}}
\end{figure}

To enlighten this problem, we have calculated the constant-volume lattice specific heat from the phonon DOS $F(\omega)$, using the conventional formula:
\begin{equation}\label{eq:heat}
C_V = R\int_0^{\infty}F(\omega)\left(\frac{\hbar\omega}{k_BT}\right)^2\frac{\exp(\frac{\hbar\omega}{k_BT})}{\left(\exp(\frac{\hbar\omega}{k_BT})-1\right)^2} d\omega.
\end{equation}
It is presented in Fig.~\ref{fig_phheat}(left panel) in the temperature range 0 - 300~K. Figure~\ref{fig_phheat}(right panel) shows $C_V/T$ {\it versus} $T^2$ in the $0-5$~K range. The points are the calculated values, and the dashed line is a fit to the Debye formula $C_V/T = \beta T^2 + \delta T^4$. 
It gives $\beta = 1.226$ (mJ mol$^{-1}$ K$^{-4}$), $\delta = 0.216$ (mJ mol$^{-1}$ K$^{-6}$), which gives $\theta_D = \left(\frac{12\pi^4 n R}{5\beta}\right)^{1/3} = 185$~K ($n = 4$). 
However, by changing the temperature range for the power-law fit, as we did for the experimental data, from 0 K to $T_{\rm max}$, we get different $\theta_D$ values, as shown in Fig.~\ref{fig_debye}, where $\theta_D$ critically depends on $T_{\rm max}$. 
Note that the power-law behavior here does not work in the temperature range below 1/50 of $\theta_D$, thus in the range that is generally considered applicable even for the $C_V = \beta T^3$ approximation.
To visualize the inaccuracy of the power-law formula with the temperature-independent $\beta$ and $\delta$ parameters even at low temperatures, in Figs.~\ref{fig_heat3}, \ref{fig_debye} we plot the "exact" calculated specific heat [eq.(\ref{eq:heat}), points] versus the power-law curves with fitting parameters obtained in different temperature ranges. For example, in the right panel of Fig.~\ref{fig_heat3} we can see that the fitting parameters from the 0 - 4 K range completely fail to describe the specific heat in 0 - 2 K.

% the experimental values (Bag {\it et al.}, Ref. [37] in the manuscript) are $\beta = 1.33$ (mJ mol$^{-1}$ K$^{-4}$), $\delta = 0.17$ (mJ mol$^{-1}$ K$^{-6}$). The $\beta$ values correspond to the Debye temperatures the experimental value is 180~K. 

This behavior is a consequence of the presence of the low-frequency softened acoustic phonon mode, which is discussed in our work. 
Due to its low frequency of about 0.5 THz in the flat part between $\Gamma$ and $X$ (see Fig. 5 in the main text), it dominates the specific heat at low temperatures. As it is not linear with respect to the wave vector, as the Debye approximation assumes, it is not surprising that specific heat cannot be described by the one- or two-parameter Debye model. Generally, $\theta_D$ is not a well-defined quantity in ScAu$_2$Al.

This shows the need for repeat measurements of the specific heat of ScAu$_2$Al in the temperature range of 0.5 - 2 K to obtain a reliable value of the Sommerfeld parameter $\gamma$.
Figure~\ref{fig_debye} (left panel) shows that below 1.5 K the power-law curves describing the lattice contribution to the specific heat start to converge, in this temperature range a precise determination of $\gamma$ is possible, and this is probably a maximal temperature for which the Debye model may be used here.

\newpage
\subsection{Cold smearing}

Cold smearing method uses the function (see Refs. [72,73]):
\begin{equation}
    \tilde{\delta}(x) = \frac{1}{\sqrt{\pi}} e^{- \left[ x -1/\sqrt{2} \right]^2}(2-\sqrt{2}x), 
\end{equation}
where $x\equiv x_{n\mathbf{k}} = (E_F-E_{n\mathbf{k}})/\sigma$, $E_{n\mathbf{k}}$ is the energy eigenvalue for given $\mathbf{k}$ and band $n$, and a broadening $\sigma=0.005$ Ry was chosen.
The average superconducting gap over the Fermi surface is computed by summing all the values of $\Delta_{n\mathbf{k}}$ (which are also computed for energies different from $E_F$) multiplied by weights, computed using the above formula, and normalized by the sum of weights:

\begin{equation}
    \overline{\Delta}_n = \frac{\sum_{\mathbf{k}} \Delta_{n\mathbf{k}}\tilde\delta(x_{n\mathbf{k}})}{\sum_{\mathbf{k}}{\tilde\delta(x_{n\mathbf{k}})}}.
\end{equation}

% \begin{figure}[H]
% 	\centering
% 	\includegraphics[width=0.45\columnwidth]{FigS3.pdf}
% 	\caption{$N^{mv}_{\mathbf{k}}$ weights from the Marzari-Vanderbilt-DeVitaPayne smearing. \label{fig_dow}}
% \end{figure}

% \newpage
\subsection{Animation of atomic vibrations}

Atomic movement is visualized using the formula for a displacement vector $\mathbf{u}_{\mathbf{q},\nu}$:
\begin{equation}
    \mathbf{u}_{\mathbf{q},\nu} = \Re \mathbf{\gamma}_{\mathbf{q},\nu} \cos(\omega_{\mathbf{q},\nu} t) + \Im \mathbf{\gamma}_{\mathbf{q},\nu} \sin(\omega_{\mathbf{q},\nu} t),
\end{equation}
where $\mathbf{\gamma}_{\mathbf{q},\nu}$ is a complex phonon eigenvector. Atoms are colored in red, yellow and blue for Sc, Au and Al, respectively.
Supplemental Material contains animations of selected acoustic phonon modes $\nu$ at $\mathbf{q}$-points in the files listed below: 
\begin{itemize}
    \item video{\textunderscore}X1.avi, video{\textunderscore}X2.avi - $\mathbf{q}=(0,1,0)$, $\nu=\{1,2\}$, w/o SOC; atoms move along base diagonal
    \item video{\textunderscore}X1so.avi, video{\textunderscore}X2so.avi - $\mathbf{q}=(0,1,0)$, $\nu=\{1,2\}$, with SOC; due to SOC the atoms change the direction of movement and do not vibrate along one line, which lowers the vibration frequency
    \item video{\textunderscore}X3.avi, video{\textunderscore}X3so.avi - $\mathbf{q}=(0,1,0)$, $\nu$=3, only Au atoms move, SOC has no effect
    \item video{\textunderscore}05XG1.avi, video{\textunderscore}05XG2.avi, video{\textunderscore}05XG3.avi - $\mathbf{q}=(0,\frac{1}{2},0)$, $\nu=\{1,2,3\}$, w/o SOC; mode $\nu$=3 is chiral 
    \item video{\textunderscore}05XG1so.avi, video{\textunderscore}05XG2so.avi, video{\textunderscore}05XG3so.avi - $\mathbf{q}=(0,\frac{1}{2},0)$, $\nu=\{1,2,3\}$, with SOC; modes are chiral
    \item video{\textunderscore}L1.avi, video{\textunderscore}L2.avi, video{\textunderscore}L3.avi, video{\textunderscore}L1so.avi, video{\textunderscore}L2so.avi, video{\textunderscore}L3so.avi - $\mathbf{q}=(\frac{1}{2},\frac{1}{2},\frac{1}{2})$, $\nu=\{1,2,3\}$
    \item video{\textunderscore}W1.avi, video{\textunderscore}W2.avi, video{\textunderscore}W1so.avi, video{\textunderscore}W2so.avi - $\mathbf{q}=(\frac{1}{2},1,0)$, $\nu=\{1,2\}$; modes are chiral
    \item video{\textunderscore}W3.avi, video{\textunderscore}W3so.avi - $\mathbf{q}=(\frac{1}{2},1,0)$, $\nu=3$
    \item video{\textunderscore}K1.avi, video{\textunderscore}K1so.avi - $\mathbf{q}=(\frac{3}{4},\frac{3}{4},0)$, $\nu=1$
    \item video{\textunderscore}K2.avi, video{\textunderscore}K3.avi, video{\textunderscore}K2so.avi, video{\textunderscore}K3.avi - $\mathbf{q}=(\frac{3}{4},\frac{3}{4},0)$, $\nu=\{2,3\}$; modes are chiral
\end{itemize}
In case of circular (chiral) phonon modes, due to the time-reversal and inversion symmetries of this full Heusler structure, the overall phonon angular momentum, carried by individual chiral modes, vanishes. However, if a related non-centrosymmetric half-Heusler structure (ScAuAl) could be synthesized and magnetically doped, then the chirality of the phonons would be worth investigating.

\end{document}